\documentclass[11pt,english,twoside]{article}

\usepackage[T1]{fontenc}
\usepackage[latin1]{inputenc}
\usepackage[english]{babel}
\usepackage{lmodern}
\usepackage{a4wide}
\usepackage{amssymb, amsmath, amsthm}
\usepackage{slashed}
\usepackage{float}
\usepackage{graphicx}
\usepackage{psfrag}
\usepackage{lscape}
\usepackage[all]{xy}
\usepackage{hyperref}
\usepackage{enumerate}
\usepackage{dsfont}
\usepackage{cite}
\usepackage{color}
\usepackage{upgreek}

\newcommand{\beq}{\begin{equation}}
\newcommand{\eeq}{\end{equation}}
\def\bea#1\eea{\begin{align}#1\end{align}}
\newcommand{\nn}{\nonumber}

\def\del {\partial}
\def\d {{\rm d}}
\def\R {\mathcal{R}}

\makeatletter
\@addtoreset{equation}{section}
\makeatother

\begin{document}

\begin{titlepage}

\rightline{\small LMU-ASC 11/13}
\rightline{\small MPP-2013-52}

\vskip 1.1cm

{\fontsize{18.2}{21}\selectfont
  \centerline{
  \textbf{Non-commutative/non-associative IIA (IIB)}}
  \vskip0.3cm
  \centerline{
  \textbf{geometries from 
  $Q$- and $R$-branes and their intersections} }}
\vskip 0.2cm
\vskip 0.9cm

\noindent\centerline{\textbf{\Large Falk Ha{\ss}ler and Dieter L\"ust}}

\vskip 0.6cm
\noindent
\textit{Arnold-Sommerfeld-Center for Theoretical Physics\\Fakult\"at f\"ur Physik, Ludwig-Maximilians-Universit\"at M\"unchen\\Theresienstra\ss e 37, 80333 M\"unchen, Germany}\\
\vskip 0.1cm\noindent
 \textit{Max-Planck-Institut f\"ur Physik\\F\"ohringer Ring 6, 80805 M\"unchen, Germany}

\vskip 0.2cm
%
\noindent {\small{\texttt{f.hassler@lmu.de, dieter.luest@lmu.de}}}

\vskip 1.1cm

\begin{center}
{\bf Abstract}
\end{center}

\noindent
In this paper we discuss the construction of non-geometric 
$Q$- and $R$-branes as sources
of non-geometric $Q$- and $R$-fluxes in string compactifications. The non-geometric $Q$-branes,
being obtained
via T-duality  from the NS 5-brane or respectively from the KK-monopole,
 are still local solutions of the standard NS action,
where however the background fields $G$ and $B$ possess non-geometric global monodromy properties. We show that using 
double field theory and redefined
background fields $\tilde G$ and $\beta$ as well as their corresponding effective action, the $Q$-branes are locally and globally well behaved solutions. Furthermore
the $R$-brane solution can be at least formally constructed using dual coordinates. We derive the associated non-geometric $Q$- and $R$-fluxes and
discuss that closed 
strings moving in the space transversal to the world-volumes of the non-geometric branes see a non-commutative or a non-associative geometry.
\\

\noindent
In the second part of the paper we construct intersecting $Q$- and $R$-brane configurations as completely supersymmetric solutions of type IIA/B supergravity
with certain $SU(3)\times SU(3)$ group structures. In the near horizon limit the intersecting brane configurations lead to type II backgrounds of the form
$AdS_4\times M_6$, where the six-dimensional compact space $M_6$ is a torus fibration with various non-geometric $Q$- and $R$-fluxes in the compact directions. It exhibits an
interesting non-commutative and non-associate geometric structure. Furthermore we also determine some of the effective four-dimensional superpotentials
originating from the non-geometric fluxes.

\vfill

\end{titlepage}

\tableofcontents

\newpage

\section{Introduction}

Non-geometric string backgrounds are interesting because of several reasons.
Already early example for classes of non-geometric string constructions, such as
covariant lattices \cite{Lerche:1986cx}, fermionic string constructions \cite{Kawai:1986va,Antoniadis:1986rn}
or asymmetric orbifolds \cite{Narain:1986qm}, have shown
that non-geometric string backgrounds are abundant and provide generic points in the string landscape.
More recently it became clear that generalized complex geometries, which include T-folds (spaces that are
globally not well-defined) \cite{Hellerman:2002ax,Dabholkar:2002sy,Shelton:2005cf,Hull:2004in} and other non-geometric string backgrounds, 
have very interesting mathematical properties. They generalize Calabi-Yau manifolds to spaces with 
several kind of geometric and non-geometric fluxes and generalized $SU(3)\times SU(3)$ group structures.
Finally, non-geometric string backgrounds naturally arise in the context of doubled field theory \cite{DFT} and T-duality,
as they can be constructed by applying chains of T-duality transformations to geometric flux backgrounds.\\

Another reason for the importance of  non-geometric string background is the observation that non-geometric fluxes are part of the 
effective 4D (super)-potential. Even without reference to an underlying non-geometric compactification from 10 dimensions, non-geometric fluxes 
already arise from the requirement of T-duality invariance of the effective scalar potential. In particular there is an interesting relation
to the gauge algebra of gauged supergravity theories and the (non)-geometric fluxes in their effective potential.
In many cases, the presence of these fluxes is important for the moduli stabilization process, and one expects to obtain phenomenologically interesting string ground
states from supergravity potentials with non-geometric fluxes.
Furthermore it was recently  shown \cite{Andriot:2011uh,Andriot:2012wx,Andriot:2012an}
that 4D non-geometric fluxes can be expressed by a new ten-dimensional effective action, where the standard $H$-flux term is
replaced by terms, which encapsulate the non-geometric fluxes. This new ten-dimensional effective action is indeed well defined
for non-geometric T-fold spaces. It can be obtained by a certain field redefinition which is closely motivated by T-duality and double field theory.\footnote{An alternative
field redefinition was discussed in \cite{Blumenhagen:2012nk,Blumenhagen:2012nt}.}\\

Finally, it was discovered that closed string coordinates in a space that is "deformed" by non-geometric fluxes, become non-commutative
and also non-associative \cite{Blumenhagen:2010hj,Lust:2010iy,Blumenhagen:2011ph,Condeescu:2012sp,Andriot:2012vb}.
The non-commutativity of closed strings in non-geometric $Q$-flux backgrounds is a non-local effect, where
the closed string commutator is proportional to the non-geometric flux times a winding number (dual momentum) 
\cite{Lust:2010iy,Condeescu:2012sp,Andriot:2012vb}:
\begin{equation}
[X_Q^i(\tau,\sigma),X_Q^j(\tau,\sigma)]  \simeq  Q^{ij}_k\, \tilde p^k\, ,
\end{equation}
with $X_Q^i(\tau,\sigma)$, $X_Q^k(\tau,\sigma)$ being the closed string coordinates in the $i,j$-directions and $\tilde p^k$ the dual momentum in the $k$-th direction.
The reason for the observed non-commutativity is that the closed string acquires mixed boundary (monodromy) conditions  (which are reminiscent of mixed D-N boundary conditions
for open strings in the presence of $F$-flux) in the presence of non-geometric fluxes of the following form:
\begin{equation}\label{monodromy}
X_Q^i(\tau,\sigma+2\pi)=  X_Q^i(\tau,\sigma)+Q^{ij}_k \,\tilde p^k ~ \tilde X_{Qj}(\tau,\sigma)\, ,
\end{equation}
where $\tilde X_{Qj}(\tau,\sigma)$ denotes the dual string coordinate in the $j$-th direction.
More generally this non-commutativity is
measured by a Wilson line operator of the $Q$-flux around holonomy circles of the non-geometric backgrounds \cite{Andriot:2012an}:
\begin{equation}\label{commcloseda}
 [ X^{i}, X^{j} ] \sim  \oint_{C_k}Q_{k}^{ij}({X})~ \d{X}^{k} \ .
\end{equation}
For so-called $R$-fluxes, which do not possess a description in terms of a local background, the closed string commutator is proportional to 
the momentum of the string:
\begin{equation}
[X_R^{i}(\tau,\sigma),X_R^{j}(\tau,\sigma)]  \simeq  R^{ijk}  \,p_k\, .
\end{equation}
Here $X_R^{i}(\tau,\sigma)$ denotes the closed string coordinate in the $R$-flux background, and the $p_k$ are the ordinary momenta.
This then also leads to the closed string  non-associativity in terms of a non-vanishing 3-bracket in  the presence of $R$-fluxes:
\begin{equation}
 \lbrack X_R^i(\tau,\sigma),X_R^j(\tau,\sigma),X_R^k(\tau,\sigma)\rbrack \simeq R^{ijk}\,.
 \end{equation}
These closed string commutation and 3-bracket  relations also  indicate a very interesting and new phase space structure, which can be also derived and quantized
using membrane sigma models \cite{Mylonas:2012pg}.\\

\vskip0.4cm
\noindent {\underline {Intersecting $Q$- and $R$-branes:}}

\vskip0.3cm
As it is well know, NS 5-branes  are supersymmetric solutions of the standard NS effective
action of type IIA/B supergravity. They
act as microscopic brane sources for the $H$-fluxes. Their T-dual configurations are the Kaluza-Klein monopoles, which are the
sources for the geometric $f$-fluxes. Hence it is natural to ask, are there also microscopic sources for the non-geometric $Q$- and $R$-fluxes?
As we will show, these branes, which we will call {\sl $Q$- and $R$-branes} can be constructed by T-duality.\footnote{The $Q$-brane solutions have been constructed before, called higher KK respectively defect branes  \cite{LozanoTellechea:2000mc,Bergshoeff:2011se} or exotic branes \cite{deBoer:2010ud,deBoer:2012ma}.}
More concretely, the $Q$-branes follow from one T-duality transformation acting in the direction transversal to the  Kaluza-Klein monopole configuration.
Their corresponding harmonic functions depend logarithmically on the two transverse directions (similar to $D7$-branes).
As we will discuss, the  $Q$-brane is the source for a non-geometric $Q$-flux. Hence it is also the source for closed string non-commutativity, 
as the space along the two "nut-directions" of the $Q$-brane becomes non-commutative for the closed string coordinates.
In addition we will also discuss that the $Q$-branes, being non-geometric solutions of the standard NS effective action, are at the same time also solutions of the
new effective action \cite{Andriot:2011uh,Andriot:2012wx,Andriot:2012an} for non-geometric string backgrounds. In fact, in terms of the redefined background fields,
where the metric $g_{ij}$ is replaced by a dual metric $\tilde g_{ij}$, and the $B_{ij}$-field gets replaced by a bi-vector $\beta^{ij}$, the $Q$-brane solution looks like 
an ordinary brane with a metric,
which is well-defined under coordinate transformations. \\

The $R$-branes are still more speculative. T-duality strongly suggests that these
8-dimensio\-nal object should exist, and it is conceivable that they can be constructed in more concrete terms using the doubled field theory formalism.
In fact, the $R$-brane metric also depends on the dual coordinates and hence cannot be given as a local function of original coordinates. In any
case, being the microscopic sources  for the $R$-fluxes, the coordinates along a three-dimensional subspace of their  world-volume are argued to  be non-associative.
So after three T-dualities, we are led to the following T-duality chain of brane solutions:
 \begin{equation}\label{eq:TdualityChain8}\boxed{
{\rm NS\,\, 5- brane} \stackrel{T_{1}}{\longrightarrow}{\rm  KK \,\, monopole}
\stackrel{T_{2}}{\longrightarrow} {\rm Q-brane}\stackrel{T_{3}}{\longrightarrow} {\rm R-brane}}
\end{equation}

\vskip0.5cm

The construction of $Q$- and $R$-branes is very useful and closely related to the problem of obtaining  6-dimensional non-commutative and non-associative
spaces, which provide consistent supersymmetric compactifications for
the type IIA/IIB superstring.
So far, closed string non-commutativity and non-associativity was discussed  for two classes of non-geometric string backgrounds. 
The first example is a  3-dimensional T-fold, being a torus fibration with elliptic ${\bf Z}_4$ monodromy supplemented by non-constant fluxes \cite{Lust:2010iy}.
In \cite{Condeescu:2012sp} this non-commutative background was further extended to a full
CFT construction, which describes a 6-dimensional, freely acting asymmetric orbifold.\\

Another class of 3-dimensional non-commutative and non-associative string  backgrounds 
is given  by the well-known chain of three T-duality transformations:
\vskip0.1cm
\begin{equation}
\boxed
{
H_{ijk} \stackrel{T_{i}}{\longrightarrow} f^{i}_{jk}
\stackrel{T_{j}}{\longrightarrow} Q_{k}^{ij}
\stackrel{T_{k}}{\longrightarrow} R^{ijk}\, .
}
\end{equation}
\vskip0.3cm
\noindent
Starting from a flat 3-torus with constant $H$-flux one successively gets a 3-dimensional twisted torus with geometric flux, a 3-dimensional T-fold with constant $Q$-flux and
finally a space with constant $R$-flux. The corresponding flux sources are given by NS 5-branes, KK monopoles and by $Q$- and $R$ branes.
However as they stand, these 3-dimensional spaces are not consistent, supersymmetric solutions of type IIA/IIB superstring theory.
Simple products of two such spaces also do not lead to consistent 6-dimensional backgrounds.
In order to generalize
 this chain of 3-dimensional spaces to consistent, 
  6-dimensional, supersymmetric solutions of type IIA/IIB supergravity, we will utilize intersecting NS 5-branes, intersecting KK monopoles,
intersecting $Q$-branes and intersecting $R$-branes. In particular we will argue that {\sl intersecting}  $Q$- and $R$-branes make physically perfect sense and lead to supersymmetric
  ground states.  We will discuss various intersecting $Q$-brane configurations and their corresponding non-vanishing closed string commutators.
 We will also discuss several intersecting $R$-brane configurations and their related 3-brackets.\\

In the near horizon limit of all these intersecting brane configurations
the 10-dimensional supersymmetric geometries will be always of the form\footnote{The geometric spaces $M_6^{H,f}$  were already constructed in \cite{Kounnas:2007dd,Caviezel:2008ik}.}
\begin{equation}
M_{10}=AdS_4\times M_6^{H,f,Q,R}\, .
\end{equation}
As we will discuss, the allowed
internal 6-dimensional spaces $M_6^{H,f,Q,R}$ are equipped with  $H,f,Q,R$-fluxes and can be derived from chains of consecutive T-dualities as follows:
\vskip0.1cm
\begin{eqnarray}\label{eq:TdualityChain}
IIA:\quad&~&  M_6^{H}
\stackrel{T_{1},T_2}{\longrightarrow}  M_6^{f}
\stackrel{T_{3},T_4}{\longrightarrow} M_6^{Q}
\stackrel{T_{5},T_6}{\longrightarrow}M_6^{R} \, ,\nonumber\\
&~&\hskip0.3cm\stackrel{T_1}\downarrow\nonumber\\
IIB:\quad&~&  \hskip0.3cm M_6^{H,f}
\stackrel{T_{2},T_3}{\longrightarrow} M_6^{f,Q}
\stackrel{T_{4},T_5}{\longrightarrow}M_6^{Q,R} \, ,\nonumber\\
&~&\hskip0.3cm\stackrel{T_3}\downarrow\nonumber\\
IIA:\quad&~&  \hskip0.3cm M_6^{H,f,Q}
\end{eqnarray}
\vskip0.3cm
\noindent
These six-dimensional spaces must possess specific $SU(3)\times SU(3)$ group structures in order to satisfy the type IIA/B supersymmetry conditions.
Specifically, $M_6^{H}$ is a flat 3-torus with $H$-fluxes in four different directions, and $M_6^{f}$ a geometric space with four different geometric fluxes.
This geometric space is just the  Nilmanifold $N_{4.7}$, which is known to provide a  
nice example for supersymmetric non-Calabi-Yau compactification with $SU(3)$ group structure.
$M_6^{Q}$ and $M_6^{R}$ will be new non-geometrical spaces with particular $SU(3)\times SU(3)$ group structures, which arise from the intersection of four $Q$-branes or, respectively, with four $R$-branes.
In addition there are also  several other allowed spaces  with mixed geometrical and non-geometrical fluxes, such as backgrounds with $H$-, $f$- and $Q$-fluxes. 
In all the  cases that involve intersections of $Q$- and/or $R$-branes one obtains an interesting pattern of different commutators and/or
3-brackets.\\


The paper is organized as follows.
In the next section we want to recall  the construction of the NS 5-branes and KK monopoles as solutions of the standard NS effective action.
Then we will move on via T-duality to the construction of the non-geometric $Q$- and $R$-branes, where we will show that these branes are good solutions
of the new effective NS action for non-geometric backgrounds. This also allows a simple derivation of the corresponding $Q$- and $R$-fluxes caused by these brane solutions.
In section three we continue  to the  configurations of four intersecting branes. Taking the near horizon limit and performing a suitable rescaling of the coordinates, $AdS_4\times M_6^{H,f,Q,R}$ geometries are derived. As we will see,
 the study of the intersecting $Q$-and $R$-branes provides a simple and elegant  way to derive all non-geometric flux backgrounds and also the  commutation relations of the internal coordinates.
 We will also provide a brief discussion about the form of the supersymmetry conditions for the intersecting non-geometric branes. In fact, using the redefined background fields
 $\tilde G$, $\beta$, $Q$ and $R$, the supersymmetry conditions can be written in a very short form, in analogy to the supersymmetry conditions for spaces with non-vanishing
 $H$-field background.
 In addition will also briefly discuss
 the effective four-dimensional superpotentials, which follow from the compactification
 on the considered geometric as well as non-geometric spaces. Specifically,  
 these compactfications will lead to effective IIA/IIB flux superpotentials \cite{Derendinger:2004jn,Villadoro:2005cu,DeWolfe:2005uu,Camara:2005dc,Shelton:2005cf}, 
 which depend on the dilaton $S$,
the K\"ahler moduli $T_i$ and the complex structure moduli $U_m$. 
 We will derive  the moduli dependence of the geometric as well as non-geometric IIA flux superpotentials.\\

\section{Geometric and non-geometric NS brane solutions}\label{sec:2}

\subsection{Geometric brane solutions}

 \vskip0.3cm
In the following, we will first briefly recall the NS 5-brane solution and the T-dual Kaluza-Klein monopole.

\subsubsection{The NS 5-brane}

Let us start with the standard effective action of type IIA/B superstrings, where we include only the NS background fields, namely the metric $G$,
the antisymmetric tensor field $B$,  its associated 3-form field $H=dB$ and the dilaton $\phi$:
\begin{equation}\label{standardNS}
  S=\int\mathrm{d}^{10}x\,  e^{-2\phi} \sqrt{|G|} \left(\R + 4(\partial \phi)^2 - \frac{1}{12} H_{ijk} H^{ijk} \right) . 
\end{equation}

As it is very well known, the NS 5-brane is a solution of the field equations of the NS effective action (\ref{standardNS}). It acts as the source for the 3-form $H$-field flux. In the string frame, the NS 5-brane is described by the following metric, anti-symmetric tensor field and dilaton configuration:
\begin{eqnarray}\label{NS5brane}
ds_{NS5}^2&=&\sum_i(dx_{\parallel}^i)^2+h(r)
\sum_k (dx_\perp^k)^2 \quad(i=0,\dots ,5\,\,{\rm and}\,\,k=1,\cdots ,4)\nonumber\\
e^\phi&=&\sqrt {h(r)}\nonumber\\
H_{mnp}&=&\epsilon_{mnpq}\partial_qh(r)
\end{eqnarray}
with the harmonic function $h$ given as $h(r) 
=1+{H\over r^2}$ ($r^2=\sum x_\perp^2$).

In order to make contact with the $H$-flux backgrounds, which will be discussed in  section three, we assume that 
the NS 5-brane is wrapping three internal, compact directions, e.g. $y^4,y^5,y^6$, and it
forms a domain wall in the four-dimensional uncompactified space-time, where we
denote the four uncompactified coordinates by $x^\mu$ ($\mu=0,\dots , 3$) and the six compact ones by $y^i$ ($i=1,\dots , 6$).
The metric in the six internal directions then takes the form
\begin{equation}\label{internal metric}
ds_{NS5}^2=h(r)\sum_{i=1,2,3} (dy^i)^2
+\sum_{i=4,5,6} (dy^i)^2\, .
\end{equation}
The corresponding 5-brane geometry is depicted in the following table:
\bigskip
\begin{center}
  \begin{tabular}{|c||c|c|c|c||c|c|c|c|c|c|}
    \hline
    & $x^{0}$ & $x^{1}$  & $x^2$  & $x^3$ & $y^1$  & $y^2$  & $y^3$  & $y^4$  & $y^5$  & $y^6$ \\
    \hline
    \hline
    $\mathrm{NS}5$ & $\bigotimes$ & $\bigotimes$  & $\bigotimes$ &   &    &   &   &$\bigotimes$
                 & $\bigotimes$  & $\bigotimes$\\
\hline
   \end{tabular}
\end{center}
\bigskip
The four-dimensional domain wall structure will be always valid in all brane configurations to be discussed in the following.
In section three we will  consider the case that the $H$-field has only legs in the transversal compact space, i.e. $H_{y^1,y^2,y^3}=H$. 
Furthermore we will consider the intersecting of four different branes, such that there remains only one common transversal direction, denoted by $x^3$.
This will be achieved by assuming that the harmonic function $h(r)$ linearly depends only on the radial direction of the four-dimensional domain wall,
associated with the coordinate $x^3$. Thus in this case
\begin{equation}\label{h const Hflux}
  h(x^3) = H x^3
\end{equation}
and after rescaling of the coordinates,
the internal six-dimensional part of the metric eq.(\ref{internal metric}) will become the flat metric of $T^6$.

\subsubsection{The Kaluza-Klein monopole}

Now we take the background eq.(\ref{NS5brane}) of the NS 5-brane and perform a T-duality transformation along the transversal compact direction $y^1\equiv y$.
As it is well-know, the T-dual configuration is given by the Kaluza-Klein monopole. \footnote{The T-duality between the NS 5-brane and the KK monopole was discussed using the double geometry formalism in \cite{Jensen:2011jna} .}
 It is a purely geometrical configuration without $H$-field and dilaton $\phi$, whose metric can be brought into the following form:
\begin{eqnarray}\label{fmetric}
ds_{KK}^2=\sum_{\mu=0,1,2}(dx^\mu)^2+\sum_{i=4,5,6} (dy^i)^2+{1\over h(r) }
\biggl(dy+\sum_{i=2,3}A_idy^i\biggr)^2
+h(r)
\biggl((dx^3)^2+
\sum_{i=2,3} (dy^i)^2\biggr).
\end{eqnarray}
Here $A_idy^i$ is a one-form gauge field that corresponds to the off-diagonal metric component of the KK monopole.
The direction $y$ is now an isometry of the solution, as the harmonic function $ h (r)
=1+{f\over r} $, $r^2=(x^3)^2+(y^2)^2+(y^3)^2$ does not anymore depend on $y$.
However this solution does not correspond to a real six-dimensional brane, but the $y$-direction is referred to be the nut direction of the KK monopole.
The corresponding nut charge is given by the parameter $f$ of the harmonic function. T-duality with the NS 5-brane implies the connection $A_i=B_{y,y^i}$ between the one-form gauge field of the KK monopole and the Kalb-Ramond field of the NS 5-brane. There is always a gauge of the Kalb-Ramond field in which $B_{y,y^3}=A_3$ vanishes. The remaining component $A_2$ is connected to the harmonic function $h$ by
\begin{equation}\label{A2 condition}
A_2=\int \mathrm{d} y^3 \, \partial_{x^3} h
\, .
\end{equation}
The KK monopole configuration is shown in the following table, with the dot denoting the nut direction:
 \bigskip
\begin{center}
  \begin{tabular}{|c||c|c|c|c||c|c|c|c|c|c|}
    \hline
    & $x^{0}$ & $x^{1}$  & $x^{2}$ & $x^{3}$ & $y$  & $y^{2}$  & $y^{3}$  & $y^4$  & $y^{5}$  & $y^{6}$ \\
    \hline
    \hline
    $\mathrm{KK}$ & $\bigotimes$ & $\bigotimes$  & $\bigotimes$ &   &  $\bullet$  &   &   &$\bigotimes$
                 & $\bigotimes$  &$\bigotimes$ \\
\hline
   \end{tabular}
\end{center}
\bigskip
Setting the harmonic function $h$ according to  eq.(\ref{h const Hflux}) and identifying $f\equiv H$r the metric in the internal compact directions will take the form:
\begin{eqnarray}
ds_{KKint}^2=
{1 \over fx^3} \biggl(dy+fy^3dy^2\biggr)^2 + fx^3 \sum_{i=2,\dots , 6} (dy^i)^2
\, .
\end{eqnarray}
It is not difficult to see that for a constant $x^3$ this metric is identical to the metric of the simplest 6-dimensional Nilmanifold, namely $N_{5.2}$.
The corresponding metric flux can be immediately read off and is given by the following non-vanishing flux component: 
\begin{equation}
f_{23}^y= f\,.
\end{equation}

\subsection{Non-geometric brane solutions}

 \subsubsection{Effective actions for non-geometric backgrounds}

Non-geometric backgrounds can be nicely described via the frame work of doubled field theory (DFT).
DFT was introduced in \cite{DFT}. 
In this theory, T-duality is turned into a manifest symmetry by doubling the coordinates at the level of the effective space-time action for string theory. T-duality relates momentum and winding modes of a closed string moving on a torus $T^D$ via the T-duality group $O(D,D)$.  When the coordinates are doubled, this duality symmetry can be made manifest. 
Thus, in DFT every conventional coordinate $x^i$, associated to momentum modes, is complemented by a dual coordinate $\tilde{x}_i$, associated to winding modes. The coordinates combine into a fundamental $O(D,D)$ vector $X^{M}=(\tilde{x}_i,x^{i})$. 
As explained in \cite{Andriot:2011uh,Andriot:2012wx,Andriot:2012an}, 
we now like to consider the following field redefinition of the metric $G$ and the $B$-field:
\begin{equation}
\label{eq:relation}
(\tilde G^{-1} + \beta)^{-1} \equiv \tilde{\cal E}^{-1} = {\cal E} =  G + B \ ,
\end{equation}
where we have introduced
\begin{equation}\label{fieldredef}
\tilde{\cal E}^{ij} = \tilde G^{ij} + \beta^{ij} \ .
\end{equation}
Here $\beta^{ij}$ is a bi-vector.
We also redefine $\phi$:
\begin{equation} \label{eq:dilintro}
\sqrt{|G|} e^{-2\phi} = e^{-2d} = \sqrt{|\tilde G|} e^{-2\tilde \phi} \ .
\end{equation}
The redefinitions (\ref{eq:relation}) has the form of a T-duality transformation over all $D$ coordinates along the tours $T^D$.
Now, the $Q$-flux is defined as  
 \begin{equation}
\label{eq:qflux}
  Q_{i}^{jk} = \partial_{i}\beta^{jk}\;.
 \end{equation}
For  non-geometric situations, where the metric $B$ and the $G$-field are only locally but not globally 
defined,  the $Q$-flux is nevertheless a globally well-defined object. 
However note that, being a partial derivative of a bi-vector, $Q$ is in general not a tensor. But, as shown in \cite{Andriot:2012wx,Andriot:2012an},
 the proper geometrical interpretation of $Q$ is playing the role of a connection,
 which allows us to construct a 
derivative for the dual $\tilde{x}$ coordinates that is covariant  {with respect to the $x$ diffeomorphisms}. 
In case $\beta$ is satisfying the simplifying condition 
\begin{equation}\label{simplcond}
\beta^{ij}\partial_{j} =0
\end{equation}
 when acting on arbitrary fields, 
the  $Q$-flux actually behaves like a tensor.

In \cite{Andriot:2011uh,Andriot:2012wx,Andriot:2012an}, we have proposed an effective action for (the NSNS sector of) non-geometric backgrounds, given in terms of the metric $\tilde{G}_{ij}$, the bivector $\beta^{ij}$ and the dilaton $\tilde{\phi}$. 
In case $\beta$ is satisfying the  condition eq.(\ref{simplcond}),
the effective action for $\tilde{G}_{ij},\beta^{ij},\tilde{\phi}$
takes the form \cite{Andriot:2011uh}
\beq\label{Qaction}
\tilde{S} =\int\mathrm{d}^{10}x\, \sqrt{|\tilde{G}|} e^{-2\tilde{\phi}} \left(\tilde{\cal R} + 4 (\partial\tilde{\phi})^2 -\frac{1}{4} Q^2\right)\ ,
\eeq
where $(\partial\tilde{\phi})^2$ and $Q^2$ are simply the squares contracted with $\tilde{G}$. 
Let us emphasize that this action has the same form as the standard NS action (\ref{standardNS}).
As we will see in the following, although the $Q$-branes are locally still solutions of the standard action (\ref{standardNS}), 
the action (\ref{Qaction}) is much  better suited to describe the $Q$-brane solutions than the standard NS one.

The non-geometric $R$-flux proposed in \cite{Andriot:2012wx,Andriot:2012an} (see also \cite{Dall'Agata:2007sr, Aldazabal:2011nj}) has the general form
\begin{equation}\label{Rflux}
  R^{ijk}  =  3\tilde{D}^{[i}\beta^{jk]} = 3\big(\tilde{\partial}^{[i}\beta^{jk]}+\beta^{l[i}\partial_{l}\beta^{jk]}\big) \ ,
\end{equation}
where $\tilde{\del}$ denotes the derivative with respect to the dual coordinate. If the simplifying condition eq.(\ref{simplcond}) is satisfied,
 the second term does not contribute, while the first gives
\begin{equation}\label{eq:RH}
 R^{ijk}  =   3\tilde{\partial}^{[i}\beta^{jk]}  \; .
\end{equation}
Since the $R$-flux also contains $\tilde{\partial}^i$-derivatives, one has to use the full DFT effective action that
in general  contains coordinates as well as dual coordinates:
\begin{equation}\label{Raction}
  S_{\rm DFT} \ = \ \int   d^{10}xd^{10}\tilde{x}\,\sqrt{|\tilde G|}\,e^{-2\tilde{\phi}}\Big[\tilde{\cal R} + 4 (\partial\tilde{\phi})^2-\frac{1}{4}R^2 
  +\dots\Big]
  \end{equation}  
  As we will see in the following, the $R$-brane solution will indeed depend  on the dual coordinate $\tilde x_i$ in one of the directions, but not at the same time on coordinate $x^i$
  in the same direction of the compact space.

Now we will move on to solutions of this new effective actions (\ref{Qaction}) and (\ref{Raction}): First we consider $Q$-branes, which are globally well defined solutions of (\ref{Qaction}). Afterwards we describe $R$-branes which are closely connected to (\ref{Raction}).

 \subsubsection{The non-commutative $Q$-brane configuration}

T-dualizing along a direction perpendicular to a KK-monopole will  result in a non-geometric background.
Specifically, starting from a single KK monopole shown in the previous subsection, we assume that  the metric eq.(\ref{fmetric})
does not depend on the coordinate $y^2$ (hence the KK monopole gets smeared in this direction). Now we can
perform a T-duality transformation along the
direction $y^2\equiv y'$. Using the Buscher rules \cite{Buscher:1987sk,Buscher:1987qj},
this operation leads to the following metric:\footnote{This background was already considered in 
\cite{LozanoTellechea:2000mc,Bergshoeff:2011se,deBoer:2010ud,deBoer:2012ma}.}
\begin{eqnarray}\label{qbrane}
ds_{Q}^2=\sum_{\mu=0,1,2}(dx^\mu)^2+\sum_{i=4,5,6}(dy^i)^2+{h( r  )\over h(r )^2+    A_2^2    }(dy^2+{dy'}^2)  +h (r )\biggl((dx^3)^2+
(dy^3)^2\biggr).
\end{eqnarray}
In addition there are also a non-vanishing $B$-field and a dilaton of the following form:
\begin{equation}\label{BQ}
B_{y,y'}=-{A_2\over  h(r )^2 +    A_2^2    }\, , \qquad e^\phi=\sqrt{h( r  )\over h(r )^2+    A_2^2    }\, .
\end{equation}

The metric (\ref{qbrane}) has the form of a 7-brane, but now there are two nut directions $y$ and $y'$ in the metric. The harmonic function $h$ only depends on two transversal
coordinates $x^3$ and $y^3$, and therefore we now get the logarithmic dependence
\begin{equation}
h( r )=  \ln r\, , \qquad r^2=(x^3)^2+(y^3)^2\, .
\end{equation}
on the transversal coordinates.
The logarithmic divergence of $h$ implies that this co-dimension two brane is ill-defined as a single brane. It does not lead to a finite energy solution.
However when we will consider intersecting branes in the next section,
we will obtain configurations that make physically good sense. In addition, 
since shifting the periodic coordinate $y^3$ by $2\pi$ does not correspond to a standard diffeomorphism of the background,  but acts as a T-duality transformation,
this  "7-brane" configuration is non-geometric. 
Its form is depicted in the following table, where the dots denote the two nut directions:
\bigskip
\begin{center}
  \begin{tabular}{|c||c|c|c|c||c|c|c|c|c|c|}
    \hline
    & $x^{0}$ & $x^{1}$  & $x^{2}$ & $x^{3}$ & $y$  & $y'$  & $y^{3}$  & $y^4$  & $y^{5}$  & $y^{6}$ \\
    \hline
    \hline
    $\mathrm{Q}$ & $\bigotimes$ & $\bigotimes$  & $\bigotimes$ &   &  $\bullet$  &$\bullet$   &   &$\bigotimes$
                 & $\bigotimes$  &$\bigotimes$ \\
\hline
   \end{tabular}
\end{center}
\bigskip

\noindent
For the harmonic function $h$ in eq.(\ref{h const Hflux}) and $Q\equiv f\equiv H$,
the two functions $h$ and $A_2$ become
\begin{equation}
h=Qx^3\, ,\quad A_2=Qy^3\, ,
\end{equation}
and  the metric in the internal six directions can be written as
\begin{eqnarray}\label{Qbrane}
ds_{Qint}^2=\sum_{i=4,5,6}(dy^i)^2+ Qx^3 dy^3 + {Qx^3 \over (Qx^3)^2+    (Qy^3)^2    }(dy^2+{dy'}^2)\, .
\end{eqnarray}
Assuming a constant $x^3$, this is nothing else than the metric of the non-geometric T-fold, which we like to call $N_{5.2}^Q$, since it is the T-dual to the Nilmanifold $N_{5.2}$. Thus this brane acts as the source of the non-geometric flux $Q$.
We will therefore call it a {\sl $Q$-brane}. Its metric (\ref{Qbrane}) is equipped with  an additional $B$-field, given now as:
\begin{equation}\label{QB}
B_{y,y'}=-{Qy^3\over (Qx^3)^2 +    (Qy^3)^2    }\, .
\end{equation}
%

The $Q$-brane background, which is specified by eq.(\ref{qbrane}) together with eq.(\ref{BQ}), is locally a solution of the standard NS action eq.(\ref{standardNS}) (see appendix \ref{Qbrane eom}).  
It was also shown in \cite{deBoer:2012ma} that this configuration preserves half of the type IIA/B supersymmetries. 
However it is much simpler to discuss this solution using the redefined background parameters $\tilde G$ and $\beta$.
Specifically, using the field redefinition eq.(\ref{fieldredef}), we obtain for $\tilde G$ and $\beta$:
\begin{eqnarray}\label{Qbranered}
d\tilde s_{Q}^2&=&\sum_{\mu=0,1,2}(dx^\mu)^2+{1\over h( r  )}\biggl(dy^2+{dy'}^2\biggr)+h (r )\biggl((dx^3)^2+
(dy^3)^2\biggr)+\sum_{i=4,5,6}(dy^i)^2\,  \, ,\nonumber\\
\beta_Q^{y,y'}&=&-A_2\, ,\nonumber\\
e^{\tilde\phi}&=& \frac{1}{\sqrt{h(r)}}\, .
\end{eqnarray}
Instead of an $H$-field, the redefined background possesses a non-vanishing $Q$-flux, which can be easily computed using eq.(\ref{eq:qflux}):
\begin{equation}
  Q_{3}^{y,y'} = \partial_{y^3}\beta_Q^{y,y'}=  -Q   \, ,
 \end{equation}
where the bi-vector $\beta^{y,y'}$ is satisfying the simplifying constraint (\ref{simplcond}).
In appendix~\ref{Qbrane eom} we show that this background is indeed a 
solution of the redefined effective action eq.(\ref{Qaction}).
Furthermore the redefined background eq.(\ref{Qbranered}) now behaves well-defined with respect to shifts of the periodic coordinate $y^3$ by $2\pi$.

Let us now discuss the non-commutative closed string geometry of $Q$-brane solution. Since it carries the $Q$-flux  $Q_{3}^{y,y'}$,
the directions $y$ and $y'$ possess  non-trivial monodromy properties, when going around the circle in the $y^3$
direction. This leads to the following closed string boundary conditions, which mixes the coordinates with the dual coordinates in the $y$ and $y'$ directions of the closed
string \cite{Andriot:2012vb}:
 \begin{eqnarray}\label{monodromy1}
Y(\tau,\sigma+2\pi)&= & Y(\tau,\sigma)+Q^{y,y'}_3 \,\tilde p^3 ~ \tilde Y'(\tau,\sigma)\, ,\nonumber\\
Y'(\tau,\sigma+2\pi)&=&  Y'(\tau,\sigma)-Q^{y,y'}_3 \,\tilde p^3 ~ \tilde Y(\tau,\sigma)\, .
\end{eqnarray}
It follows that
a closed string in the field of the $Q$-brane sees a non-commutative geometry in the $y,y'$-directions:
\begin{equation}
[Y(\tau,\sigma),Y'(\tau,\sigma)]\simeq Q~ \tilde  p^3\, .
\end{equation}
Here $\tilde p_3$ is the dual momentum in the $y^3$-direction.

\subsubsection{The non-associative  $R$-brane configuration}

Let us now come to the final step in the T-duality chain eq.(\ref{eq:TdualityChain}). 
Starting from the $Q$-brane,  we will assume that the function $h( r  )$ does not depend anymore on
the coordinate $y^3$. Hence $h( r  )$ will be a linear function in the remaining transversal coordinate $x^3$. Then the T-duality in $y^3=y''$ leads to the following 8-dimensional $R$-brane configuration:
\bigskip
\begin{center}
  \begin{tabular}{|c||c|c|c|c||c|c|c|c|c|c|}
    \hline
    & $x^{0}$ & $x^{1}$  & $x^{2}$ & $x^{3}$ & $y$  & $y'$  & $y''$  & $y^4$  & $y^{5}$  & $y^{6}$ \\
    \hline
    \hline
    $\mathrm{R}$ & $\bigotimes$ & $\bigotimes$  & $\bigotimes$ &   &  $\bullet$  &$\bullet$   & $\bullet$  &$\bigotimes$
                 & $\bigotimes$  &$\bigotimes$ \\
\hline
   \end{tabular}
\end{center}
\bigskip

This brane configuration apparently possess three nut directions.
However, since we are now doing a T-duality along a non-isometry direction, namely $y''$, the $R$-brane does not possess a local metric in the original coordinates. 
But, as discussed in \cite{Andriot:2012an}, double field theory \cite{DFT} has a proposal on how to T-dualise along a direction which is not an isometry: we just need here to formally replace the coordinate $y''$ by its dual coordinate $\tilde y''$, in analogy to the replacement of the momentum by its dual quantity, namely the winding. 
Performing this replacement, we get using the redefined background fields  $\tilde G$ and $\beta$ the following expressions:
\begin{eqnarray}\label{Rbranered}
d\tilde s_{R}^2&=&\sum_{\mu=0,1,2}(dx^\mu)^2+{1\over h(r)}\biggl(dy^2+{dy'}^2\biggr)+h(r)\biggl((dx^3)^2+{d\tilde{y}''}^2\biggr) + \sum_{i=4,5,6}(dy^i)^2\,  \, ,\nonumber\\
\beta_R^{y,y'}&=&-R\tilde y''\, ,\nonumber\\
e^{\tilde\phi}&=&{1\over \sqrt {h(r)}} \, ,
\end{eqnarray}
where after T-duality we have denoted the parameter of the solution by $R$, with  $R\equiv Q\equiv f\equiv H$.
Using eq.(\ref{eq:RH}), the corresponding $R$-flux is given as
\begin{equation}
  R^{y,y',y''} = \partial_{\tilde y''}\beta_R^{y,y'}=  -R  \,.
 \end{equation}

A closed string in the field of the $R$-brane sees a non-commutative and non-associative geometry in the $y,y',y''$-directions.
Specifically we obtain 
 the following non-vanishing commutators and 3-brackets with $p,p'p''$ being  the momenta in the $y,y',y''$-directions:
\begin{eqnarray}
\lbrack Y(\tau,\sigma),Y'(\tau,\sigma)\rbrack&\simeq&R ~p''\nonumber \\
\lbrack Y''(\tau,\sigma),Y(\tau,\sigma)\rbrack&\simeq&R~ p'\nonumber \\
\lbrack Y'(\tau,\sigma),Y''(\tau,\sigma)\rbrack&\simeq&R ~p\nonumber \\
\lbrack Y(\tau,\sigma),Y'(\tau,\sigma),Y''(\tau,\sigma)\rbrack &\simeq &R\, .
\end{eqnarray}
\\

\section{Type IIA/B $AdS_4\times M_6$ backgrounds from intersecting NS 5-branes, KK monopoles, $Q$- and $R$-branes}

\subsection{Geometric intersecting branes}

In the absence of sources and higher-order derivative corrections, 
supersymmetric backgrounds of type II supergravity 
of the form $\mathbb{R}^{1,3}\times{M}_6$ require the internal manifold 
$M_6$ to be Calabi-Yau. 
Turning on the background fluxes while preserving maximal symmetry in the four non-compact dimensions  
forces the background to be of the form of a warped product $AdS_4\times_w \mathcal{M}_6$, where
 ${M}_6$ is no longer a Calabi-Yau. 
The departure from the Calabi-Yau condition in the presence of fluxes 
can be described 
by $SU(3)$ group structures and associated torsion classes \cite{Gurrieri:2002wz,LopesCardoso:2002hd}
or more generally
by reformulating the supersymmetry conditions in the 
framework of generalized geometry \cite{hitchin, gualtieri}. This leads to 
the statement that  ${M}_6$ must possess a pair of compatible
 pure spinors obeying certain differential conditions \cite{Grana:2005sn}.

Let us very briefly recall the supersymmetry conditions on the background
$(G,B,\phi)$ for type IIA/IIB compactifications. 
We perform a four-plus-six space-time split, according to which the 
ten-dimensional metric takes the warped-product form:
\begin{equation}
ds^2=e^{2A(y)}ds^2_4+g_{ij}dy^idy^j
~,
\end{equation}
where $\exp\!( 2 A(y) )$ is the warp factor, $ds^2_4$ is the line element of $AdS_4$ and $g_{ij}$ is the internal-manifold metric. 
(We denote the four coordinates of $AdS_4$ by $x^\mu$ ($\mu=0,\dots , 3$), whereas
the six compact coordinates are denoted by $y^i$ ($i=1,\dots , 6$).)
Within the framework of generalized complex geometry, supersymmetry requires
that the structure group of the {\em generalized} tangent bundle to be SU(3)$\times$SU(3). 
The supersymmetry generators $\eta^{(1)}$ and $\eta^{(2)}$ 
are then collected into two spinor bilinears, which can be associated with two polyforms of definite degree
\begin{equation}\label{polys}
\slash\hskip-0.2cm
{\Psi}_+ = \frac{8}{|a||b|}\eta^{(1)}_+ \otimes \eta^{(2)\dagger}_+ \, , \qquad
\slash\hskip-0.2cm
{\Psi}_- = \frac{8}{|a||b|}\eta^{(1)}_+ \otimes \eta^{(2)\dagger}_- \, .
\end{equation}
In order to obtain similar equations in IIA and IIB one redefines
\begin{equation}\label{A/B}
\Psi_1 = \Psi_{\mp} \, , \qquad \Psi_2 = \Psi_{\pm} \, ,
\end{equation}
with upper/lower sign for IIA/IIB. With these definitions the supersymmetry conditions take the following schematic form in both
IIB and IIA, where we neglected all possible Ramond fluxes as well as a warp factor and the dilaton
\begin{eqnarray}
\label{gensusy}
d_H   \Psi_1  &=& 0\, ,\nonumber\\
 d_H   \Psi_2  &=& 0\, ,\end{eqnarray}
where
\begin{equation}
d_H=d+H\wedge\, .
\end{equation}

\vskip0.4cm
\noindent {\underline {$AdS_4$ vacua in Type IIA:}}

\vskip0.3cm

\noindent

Necessary and sufficient conditions for supersymmetric vacua   of the form $AdS_4\times M_6$
have been established in the case of
  (massive) IIA backgrounds with  constant 
dilaton and warp factor  in \cite{Lust:2004ig}. 
First we  discuss IIA solutions with strict $SU(3)$ group structure for the cases of
geometric spaces with $H$-fluxes and/or geometric fluxes, 
where we  mainly follow the discussion in \cite{Lust:2004ig,Caviezel:2008ik}.
Neglecting the Ramond forms, the solutions are given by\footnote{The parameter $H$ is proportional to the mass parameter $m$ of massive IIA supergravity $H={2\over 5}m$.  Furthermore $m$ is related to 
$W^{-1}$, the radius of $AdS_4$, in the following way:
$
W=-{1\over 5} e^\phi m$. In general,
this equation also contains the Freund-Rubin parameter.}
\begin{equation}
H (y)= He^{\phi}\Re \Omega \, .\\
\end{equation}
Furthermore, ($J$, $\Omega$) is the SU(3)  group structure of the internal six-manifold,
i.e.\ $J$ is a real two-form, and $\Omega$ is a decomposable complex three form such that:
\begin{eqnarray}\label{suthree}
\Omega\wedge J&=&0 \, , \nonumber\\
\Omega\wedge\Omega^*&=&\frac{4i}{3}J^3\neq 0\, .
\end{eqnarray}
The $SU(3)$ group structure is related to the two polyforms
by making the following identification:
\begin{equation}
\label{SU3pure}
\Psi_- = - \Omega \, , \qquad \Psi_+ = e^{iJ}  \, ,
\end{equation}
Supersymmetry is then requiring that
the only non-zero torsion classes of the internal manifold are ${\cal W}^-_1,{\cal W}^-_2$
which are defined such that:\footnote{Consistent supersymmetric IIA solutions  will also require the existence of Ramond 2-form and 4-form fluxes.}
\begin{eqnarray}
\label{torsionclasses}
\d J&=&-\frac{3}{2}i \, \mathcal{W}_1^- \Re \Omega\, , \nonumber\\
\d \Omega&=& \mathcal{W}^-_1 J\wedge J+\mathcal{W}^-_2 \wedge J\, .
\end{eqnarray}
This requirement largely restricts the number of known solutions. By taking the internal six-dimensional space to be a 6-dimensional
Nilmanifold ${M_6}$, it suffices to look for all possible
Nilmanifolds,
whose only non-zero torsion classes are ${\cal W}^-_{1,2}$. As we will discuss in the following, there are
exactly two possibilities in type IIA, namely the six-torus and
the Nilmanifold $N_{4.7}$ of Table 4 of \cite{Grana:2006kf} (also known as the Iwasawa manifold).

\vskip0.4cm
\noindent {\underline {$AdS_4$ vacua in Type IIB:}}

\vskip0.3cm\noindent
For IIB on the other hand, where
the above definitions of $\Psi_1$ and $\Psi_2$ are switched, we are interested in the case for which
\begin{equation}
\eta^{(2)}_+ = V^i \gamma_i \eta^{(1)}_- \, .
\end{equation}
This condition will correspond to spaces with static $SU(2)$ group structure, and we
define the following $SU(2)$-structure quantities
\begin{equation}
\omega_2 = J + 2i V \wedge V^* \, , \qquad \Omega_2 = \iota_{V^*} \Omega
\end{equation}
where $J$ and $\Omega$ form the $SU(3)$-structure associated to $\eta_+=|a|^{-1} \eta^{(1)}_+$.
We then find for the pure spinors
\begin{eqnarray}
\label{SU2pure}
\Psi_+ & =& - e^{2 V \wedge V^*} \Omega_2 \, , \nonumber\\
\Psi_- & = &- 2 \, V \wedge e^{i \omega_2} \, .
\end{eqnarray}
Plugging this ansatz in (\ref{gensusy}), one finds equations for the SU(2)-structure
quantities $V$, $\omega_2$ and $\Omega_2$.

We will now first consider six-dimensional geometric background, which arise from the intersection of NS 5-branes and KK monopoles.

\subsubsection{Type IIA: six-torus with $H$-flux}\label{six torus H}

Here the six-dimensional space $M_6$ is just the flat torus $T^6$.
Let us define a left-invariant basis $\{e^i\}$ such that:
\begin{equation}
d e^i = 0, \qquad i=1,\dots, 6~.
\end{equation}
On the torus we can just choose $e^i = \d y^i$, where $y^i$ are the internal coordinates.
The SU(3)-structure is given by
\begin{align}\label{jot}
J&=e^{12}+e^{34}+e^{56} \, ,\nn\\
\Omega&=(ie^1+e^2)\wedge(ie^3+e^4)\wedge(ie^5+e^6)~,
\end{align}
Hence all torsion
classes trivially vanish in this case. 
In addition, there are four  non-vanishing $H$-field components:
\begin{equation}\label{hsixtorus}
H_{y^2,y^4,y^6}=H_{y^2,y^5,y^3}=H_{y^1,y^6,y^3}=H_{y^1,y^5,y^4} = H\, .
\end{equation}
The corresponding $B$-field components are
\begin{equation}
B_{y^2,y^4}=-B_{y^1,y^3} =Hy^6 \, ,\quad    B_{y^1,y^4}=B_{y^2,y^3}=-Hy^5 \, .
\end{equation}

As shown in \cite{Kounnas:2007dd,Caviezel:2008ik},
this background can be obtained 
from the 
intersection of four different NS 5-branes, as shown in the following table:\footnote{In addition, the complete brane configuration also contains three intersecting
D4-branes and one D8-brane, which are the sources of the omitted Ramond 4-form fluxes and the IIA mass parameter $m$, as well as four (smeared) orientifold O6-planes.}

\bigskip
\begin{center}
\begin{tabular}{|c||c|c|c|c|c|c|c|c|c|c|}
    \hline
    & $x^0$ & $x^1$  & $x^2$  & $x^3$ & $y^1$  & $y^2$  & $y^3$  & $y^4$  & $y^5$  & $y^6$ \\
    \hline
    \hline
    \hline
    $\mathrm{NS}5$ & $\bigotimes$ & $\bigotimes$  & $\bigotimes$ &   &  $\bigotimes$  &   &  $\bigotimes$ &
                 & $\bigotimes$  & \\
\hline
    $\mathrm{NS}5^{\prime}$ & $\bigotimes$ & $\bigotimes$  & $\bigotimes$ &   &  $\bigotimes$  &  &   &  $\bigotimes$
                 &  & $\bigotimes$ \\
 \hline
    $\mathrm{NS}5^{\prime\prime}$ & $\bigotimes$ & $\bigotimes$  & $\bigotimes$ &   &  & $\bigotimes$ &   &
               $\bigotimes$    &  $\bigotimes$  & \\
\hline
    $\mathrm{NS}5^{\prime\prime\prime}$ & $\bigotimes$ & $\bigotimes$  & $\bigotimes$ &   &  & $\bigotimes$ &  $\bigotimes$     &
               &  &  $\bigotimes$ \\
 \hline
   \end{tabular}
\end{center}
\bigskip

The full intersecting brane solution patches two asymptotic regions: a
near-horizon region of  $AdS_4\times T^6$ geometry and a flat  region
at infinity. Note that the harmonic function of the intersecting branes only depend on the non-compact (radial) coordinate of the four-dimensional $AdS_4$ space, denoted by
$x^3$. As shown in \cite{Kounnas:2007dd},  
in the near-horizon limit $x^3\rightarrow 0$ of the
solution,  the brane system above is replaced by fluxes. After rescaling of the coordinates, the near-horizon limit of the
solution precisely corresponds to the flat six-torus $T^6$ with  $H$-field, as shown eq.(\ref{hsixtorus}).

Furthermore,
in type IIA (orientifold) compactifications, the effective, 4-dimensional superpotential has the following general  structure:
\begin{eqnarray}
H-{\rm flux:}\quad W_H^{IIA}=a_0S+b_mU_m\,.
\end{eqnarray}
Here we obtain from the NS sector:
\begin{eqnarray}
  W_H^{IIA}=H_{246}S+H_{235}U_1+H_{145}U_2+H_{136}U_3\, .
\end{eqnarray}
Note that the superpotential with all NS and Ramond
fluxes, which follow from the complete intersection of  NS 5-branes branes plus  the additional D-branes,
 stabilizes all moduli  in an supersymmetric $AdS_4$ minimum. In this context it is interesting to see that the values of the stabilized moduli fields in the
 effective field theory are identical to the asymptotic values of these fields in the near
 horizon limit of the intersecting brane geometries \cite{Kounnas:2007dd}. 

\subsubsection{Type IIA: Iwasawa manifold}

Let us now perform two T-duality transformations along the  $y^1$ and $y^2$ directions of the internal manifolds.
These will transform all four $H$-field components into the following geometric fluxes:
\begin{equation}\label{iwasawafluxes}
f^2_{46}=-f^1_{36}=-f^1_{45}=-f^2_{35}=f
\end{equation}
where by T-duality $f\equiv H$. 
Then the 6-dimensional metric takes the following form
\begin{eqnarray}\label{iwametric}
  ds^2&=&\sum_i \biggl(dy^i+f^i_{jk}y^kdy^j\biggr)^2 \nonumber\\
&=& \biggl(dy^1-fy^6dy^3-  fy^5dy^4   \biggr)^2
+\biggl(dy^2+fy^6dy^4-   fy^5dy^3   \biggr)^2 + \sum_{i=3,4,5,6}
(dy^i)^2 \, .\nn
\end{eqnarray}
The corresponding 6-dimensional space is just the Nilmanifold $N_{4.7}$, known as Iwasawa manifold, as can be seen as follows.
The left-invariant basis
is now defined by:
\begin{eqnarray}
\d e^a&=&0,~~a=3,\dots, 6 \, , \nonumber \\
\d e^1&=&f(e^{45}-e^{36}) \, ,\nonumber \\
\d e^2&=&f(e^{46}+e^{35})~,
\end{eqnarray}
and is usually denoted by $(45-36,46+35,0,0,0,0)$. Up to basis transformations there
is a unique SU(3)-structure satisfying the supersymmetry conditions:
\begin{eqnarray}
J & =& e^{12}+e^{34}+  e^{65} \, , \nonumber\\
\Omega&=& \, (e^1+ie^2)\wedge(e^3-ie^4)\wedge(e^5+ie^6)\,.
\label{lolut}
\end{eqnarray}
In the left-invariant basis, the metric is  simply given by $g=\text{diag} (1,1,1,1,1,1)$, and the torsion classes
can be read off from $d J$, $d \Omega$ as follows:
\begin{eqnarray}
{\cal W}^-_1 & = &-\frac{2if}{3}  , \nonumber\\
{\cal W}^-_2 & =& -\frac{4if}{3}  \, \left(2e^{12}+e^{34}+ e^{56}\right) \, ,
\label{expltord}
\end{eqnarray}
while all other torsion classes vanish. 
We can verify that $\d{\cal W}^-_2$ is
proportional to $\Re\Omega$:
\begin{align}
d{\cal W}^-_2=-\frac{8if^2}{3}\Re\Omega~.
\label{dw2d}
\end{align}

This geometric space corresponds to the near horizon geometry of the  intersection of four different
KK monopoles as shown in the following table (We do not display the non-compact directions anymore, but let us keep in mind that they
form domain walls.):
\begin{center}
  \begin{tabular}{|c|c|c|c|c|c|c|}
    \hline
    & $y^1$  & $y^2$  & $y^3$  & $y^4$  & $y^5$  & $y^6$ \\
    \hline
     \hline
       $\mathrm{KK}$ &   $\bigotimes$  & $\bullet$ &  $\bigotimes$ &
                 & $\bigotimes$ &  \\
\hline
    $\mathrm{KK}^{\prime}$    &  $\bigotimes$  & $\bullet$ &   &  $\bigotimes$
                 &  & $\bigotimes$ \\
 \hline
    $\mathrm{KK}^{\prime\prime}$    & $\bullet$ & $\bigotimes$ &   &
               $\bigotimes$    &  $\bigotimes$  & \\
\hline
    $\mathrm{KK}^{\prime\prime\prime}$    & $\bullet$ & $\bigotimes$ &  $\bigotimes$     &
               &  &  $\bigotimes$ \\
\hline
   \end{tabular}

\end{center}
\vskip0.5cm
 The IIA effective superpotential of geometric fluxes has the following general  structure:
\begin{eqnarray}
f-{\rm flux:}\quad W_f^{IIA}=a_iST_i+b_{im}T_iU_m\,.
\end{eqnarray}
Using precisely the fluxes in eq.(\ref{iwasawafluxes})
we obtain
\begin{eqnarray}
 W_f^{IIA}=f^2_{46}ST_1+f^2_{35}T_1U_1+f^1_{45}T_1U_2+f^1_{36}T_1U_3\, .
\end{eqnarray}

\subsubsection{Type  IIB: the Nilmanifold $N_{5.1}$}\label{nilmanifold N5.1}

Starting from the flat 6-torus with $H$-flux we can also perform only one T-duality along the $y^1$ direction (or from the Iwasawa manifold via T-duality along  $y^2$).
This leads us to a IIB background with the following two geometric fluxes:
\begin{equation}
-f^1_{36}=-f^1_{45}=f\,.
\end{equation}
In addition two $H$-fluxes survive the T-duality transformation:
\begin{equation}
H_{246}=-H_{235}=H  \, . 
\end{equation}
Now the metric is given by
\begin{eqnarray}
ds^2&=&\biggl(dy^1-fy^6dy^3-  fy^5dy^4   \biggr)^2 +
\sum_{i=2,3,4,5,6} (dy^i)^2\, ,\nn
\end{eqnarray}
which corresponds to the Nilmanifold $N_{5.1}$, defined left-invariant basis
\begin{eqnarray}
\d e^a&=&0,~~a=2,3,\dots, 6 \, , \nonumber \\
\d e^1&=&f(e^{45}-e^{36}) \, .\nonumber \\
\end{eqnarray}
For the $SU(2)$ group structure quantities one obtains
\begin{eqnarray}
e^{i\theta} V & = &\frac{1}{2} \left( e^1 +i e^2 \right) \, , \nonumber\\
\omega_2 & =& -e^{36} + e^{45} \, , \nonumber\\
\Omega_2 & = &-i e^{i \theta} (i e^6 +e^3)\wedge (i e^4 + e^5 ) \, .
\end{eqnarray}

Again, the above IIB background is the near horizon geometry of its brane sources, namely
 the intersection of two NS 5-branes and two KK monopoles, as shown in the following table:
\bigskip
\begin{center}
  \begin{tabular}{|c|c|c|c|c|c|c|}
    \hline
    & $y^1$  & $y^2$  & $y^3$  & $y^4$  & $y^5$  & $y^6$ \\
    \hline
     \hline
    $\mathrm{NS}5$ &   $\bigotimes$  &   &  $\bigotimes$ &
                 & $\bigotimes$ &  \\
\hline
    $\mathrm{NS}5^{\prime}$    &  $\bigotimes$  &  &   &  $\bigotimes$
                 &  & $\bigotimes$ \\
 \hline
    $\mathrm{KK}^{\prime\prime}$    & $\bullet$  & $\bigotimes$ &   &
               $\bigotimes$    &  $\bigotimes$  & \\
\hline
    $\mathrm{KK}^{\prime\prime\prime}$    & $\bullet$ & $\bigotimes$ &  $\bigotimes$     &
               &  &  $\bigotimes$ \\
\hline
   \end{tabular}

   \end{center}
\vskip0.7cm
 The chain of T-dualities for the three IIA/IIB geometrical brane intersections and their near horizon geometries obtained so far  is summarized as follows:
 \begin{equation}\label{eq:TdualityChain1}\boxed{
T^6+4H |_{IIA}\stackrel{T_{1}}{\longrightarrow} N_{5.1}+ 2H|_{IIB}
\stackrel{T_{2}}{\longrightarrow} N_{4.7}|_{IIA}}
\end{equation}

\vskip1.5cm

\subsection{Non-geometric spaces: Intersecting $Q$-branes and $R$-branes}\label{sec:3}


Let us start this section by 
briefly discussing the supersymmetry conditions on 6D spaces with non-geometric fluxes. This can be done most elegantly using the redefined background fields $\tilde G,\beta,\tilde\phi$. 
Since, under the simplifying condition (\ref{simplcond}), the old action (\ref{standardNS}) and the new action  (\ref{Qaction}) have the same form, supersymmetry still requires the structure group of the {generalized} tangent bundle to be $SU(3)\times SU(3)$. However now the supersymmetry conditions are evaluated with respect to the space determined by $\tilde G,\beta,\tilde\phi$.\footnote{Since
the field redefinition eq.(\ref{eq:relation}), which involves an exchange of 2-form $B_{ij}$ with the bi-vector $\beta^{ij}$, basically exchanges the tangent bundle with the co-tangent bundle, the two $SU(3)$ group factors also get exchanged.}
Hence we can again define  two polyforms
\begin{equation}\label{polysnew}
\slash\hskip-0.2cm
\tilde{\Psi}_+ = \frac{8}{|a||b|}\tilde\eta^{(1)}_+ \otimes \tilde\eta^{(2)\dagger}_+ \, , \qquad
\slash\hskip-0.2cm
\tilde{\Psi}_- = \frac{8}{|a||b|}\tilde\eta^{(1)}_+ \otimes \tilde \eta^{(2)\dagger}_- \, ,
\end{equation}
where $\tilde\eta^{(1,2)}$ are spinors on the redefined background space.
The supersymmetry conditions then take the following  schematic form for the case of $Q$-flux under the simplifying condition (\ref{simplcond}):\footnote{The general case
and also the case of non-vanishing $R$-flux is more involved and will be treated by D. Andriot and A. Betz.}
\begin{eqnarray}
\label{gensusynew}
d_Q  \tilde\Psi_1  &=& 0\, ,\nonumber\\
d_Q   \tilde\Psi_2 & =& 0\, ,
\end{eqnarray}
where now 
\begin{equation}
d_Q=d+Q\cdot \,.
\end{equation}
Here the $\cdot$ defines the proper contraction of the $Q$-flux on vectors and forms (see also \cite{Grana:2008yw} for a discussion about supersymmetry, T-duality and
non-geometric backgrounds).

For explicit examples the generalized $SU(3)\times SU(3)$ structure group  will be reduced to a space with $SU(3)$ or with $SU(2)$ group structure.
As we will now discuss, the IIA non-geometric background with four $Q$-fluxes will be a space with flat metric $\tilde G_{ij}$ (a six-dimensional torus) after the field redefinition.
On the other hand, the IIB space with two geometric fluxes and two $Q$-fluxes  will possess a $SU(2)$ group structure just like its T-dual IIB counterpart, namely the Nilmanifold $N_{5.1}$ equipped with two $H$-fluxes.\\
At the end of this subsection let us also discuss the non-trivial monodromy properties and the associated non-commutative  geometry structure
of the intersecting brane configurations.
All considered spaces  $M_6$ are fibrations of a four-dimensional torus $T^4$ with coordinates $y^i$ ($i=1,\dots ,4$) over a two-dimensional torus with coordinates denoted
by $y^m$ ($m=5,6$). The monodromy properties of $M_6$ are specified by the fluxes $H,f,Q$, when going around two circles in the directions $y^5$ and $y^6$. From these we can then derive
the following closed string boundary (monodromy) conditions:
\begin{eqnarray}
Y^m(\tau,\sigma+2\pi)&= & Y^m(\tau,\sigma)+2\pi ~\tilde p^m \, ,\nonumber\\
Y^i(\tau,\sigma+2\pi)&=&  Y^i(\tau,\sigma)+f^{i}_{jm}\,\tilde p^m ~  Y^j(\tau,\sigma)+Q^{ij}_m \,\tilde p^m ~ \tilde Y_j(\tau,\sigma)\, ,\nonumber\\
\tilde Y_i(\tau,\sigma+2\pi)&=&  \tilde Y_i(\tau,\sigma)-f^{j}_{im}\,\tilde p^m ~ \tilde  Y_j(\tau,\sigma)+H^{ijm} \,\tilde p^m ~  Y^j(\tau,\sigma)\, .
\end{eqnarray}
Here the $\tilde p^m$ are the winding numbers of the closed string in the $y^m$-directions. Following \cite{Andriot:2012vb}, we can derive from these closed string boundary
conditions the following non-vanishing commutators among the closed string coordinates in the directions $y^i$ ($i=1,\dots ,4$):
\begin{eqnarray}
[Y^i(\tau,\sigma),Y^j(\tau,\sigma)] & \simeq & Q^{ij}_m ~\tilde p^m\, .
\end{eqnarray}

\subsubsection{Type IIA: four $Q$-fluxes}\label{sec:fourQfluxes}

The metric and the $H$-field of four intersection Q-Branes
\bigskip
\begin{center}
 \begin{tabular}{|c|c|c|c|c|c|c|}
    \hline
    & $y^1$  & $y^2$  & $y^3$  & $y^4$  & $y^5$  & $y^6$ \\
    \hline
     \hline
       $\mathrm{Q}$ &   $\bigotimes$  & $\bullet$ &  $\bigotimes$ &$\bullet$
                 & $\bigotimes$ &  \\
\hline
    $\mathrm{Q}^{\prime}$    &  $\bigotimes$  & $\bullet$ & $\bullet$  &  $\bigotimes$
                 &  & $\bigotimes$ \\
 \hline
    $\mathrm{Q}^{\prime\prime}$    & $\bullet$ & $\bigotimes$ & $\bullet$  &
               $\bigotimes$    &  $\bigotimes$  & \\
\hline
    $\mathrm{Q}^{\prime\prime\prime}$    & $\bullet$ & $\bigotimes$ &  $\bigotimes$     &$\bullet$
               &  &  $\bigotimes$ \\
\hline
   \end{tabular}
   \end{center}
   are derived form the D4/D8/NS5 solution presented in \cite{Kounnas:2007dd}. This solution is parameterized by eight different function: $H^\text{D8}$ for the D8-brane, $H^\text{D4}_i$ ($i=1,2,3$) for the three D4-branes and $H^\text{NS5}_i$ ($i=1,2,3,4$) for the four NS 5-branes. To adapt this solution to the one introduced in section \ref{six torus H}, we use
\begin{align*}
   H_{246}&=\del_{x^3} H^\text{NS5}_1\,(H^\text{D8})^{-1} &
   H_{235}&=-\del_{x^3} H^\text{NS5}_2\,(H^\text{D8})^{-1} \\
   H_{136}&=-\del_{x^3} H^\text{NS5}_3\,(H^\text{D8})^{-1} &
   H_{145}&=-\del_{x^3} H^\text{NS5}_4\,(H^\text{D8})^{-1}
\end{align*}
which differs by the sign of $H_{246}$ from \cite{Kounnas:2007dd}.
T-dualizing in the $y^1$, $y^2$, $y^3$ and $y^4$ directions and demanding
\begin{equation*}
  H^\text{NS5}_1 H^\text{NS5}_2 \del_{x^3} H^\text{NS5}_3\,\del_{x^3} H^\text{NS5}_4 = H^\text{NS5}_3 H^\text{NS5}_4 \del_{x^3} H^\text{NS5}_1\,\del_{x^3} H^\text{NS5}_2
\end{equation*}
lead to the internal metric 
\begin{gather*}
  ds^2_Q=\left(H^\text{D8} \prod_{i=1}^3 H^\text{D4}_i\right)^{1/2} \Biggl[
    {H^\text{NS5}_1 H^\text{NS5}_2 \over H^\text{D4}_2 x}(dy^1)^2 +
    {H^\text{NS5}_3 H^\text{NS5}_4 \over H^\text{D4}_2 x}(dy^2)^2 +
    {H^\text{NS5}_1 H^\text{NS5}_4 \over H^\text{D4}_1 x}(dy^3)^2 + \\
    {H^\text{NS5}_2 H^\text{NS5}_3 \over H^\text{D4}_1 x}(dy^4)^2 +
    {H^\text{NS5}_2 H^\text{NS5}_4 \over H^\text{D4}_3 H^\text{D8}} (dy^5)^2 +  
  {H^\text{NS5}_1 H^\text{NS5}_3 \over H^\text{D4}_3 H^\text{D8}} (dy^5)^2 \Biggr] \\[0.5em]
  \text{with}\quad x=\left(\prod_{i=1}^4 H^\text{NS5}_i\right) + {1\over H^\text{D8}}\biggl( \del_{x^3}H^\text{NS5}_2\,\del_{x^3} H^\text{NS5}_4 (y^5)^2 + \del_{x^3} H^\text{NS5}_1\, \del{x^3} H^\text{NS5}_3 (y^6)^2\biggr) 
\end{gather*}
and the $B$-field
\begin{align*}
   B_{31}&= {\del_{x^3} H^\text{NS5}_1 y^6 \over x} &
   B_{32}&= {\del_{x^3} H^\text{NS5}_2 y^5 \over x} \\
   B_{42}&= {\del_{x^3} H^\text{NS5}_3 y^6 \over x} &
   B_{41}&= -{\del_{x^3} H^\text{NS5}_4 y^5 \over x}
\end{align*}
of four intersection $Q$-branes. As already stressed, it is often more convenient to use the redefined metric $\tilde G$ and the bi-vector $\beta$ for $Q$-brane configurations. After applying the corresponding transformation in eq.(\ref{fieldredef}), one gets
\begin{gather*}
  d\tilde{s}^2_Q=\left(H^\text{D8} \prod_{i=1}^3 H^\text{D4}_i\right)^{-
  1/2} \Biggl[
    H^\text{D4}_2 H^\text{D4}_3 \left(
      H^\text{NS5}_3 H^\text{NS5}_4 (dy^1)^2 +
      H^\text{NS5}_1 H^\text{NS5}_2 (dy^2)^2 \right) + \\
    H^\text{D4}_1 H^\text{D4}_3 \left(
      H^\text{NS5}_2 H^\text{NS5}_3 (dy^3)^2 +
      H^\text{NS5}_1 H^\text{NS5}_4 (dy^4)^2 \right) + \\
     H^\text{D4}_3 H^\text{D8} \left(
     {1 \over H^\text{NS5}_2 H^\text{NS5}_4} (dy^5)^2 +
     {1 \over H^\text{NS5}_1 H^\text{NS5}_3} (dy^6)^2 \right)
  \Biggr]
\end{gather*}
and
\begin{align*}
   \beta^{31}&= -{\del_{x^3} H^\text{NS5}_1 y^6 \over H^\text{D8}} &
   \beta^{32}&= -{\del_{x^3} H^\text{NS5}_2 y^5 \over H^\text{D8}} \\
   \beta^{42}&= -{\del_{x^3} H^\text{NS5}_3 y^6 \over H^\text{D8}}&
   \beta^{41}&=  {\del_{x^3} H^\text{NS5}_4 y^5 \over H^\text{D8}} \,,
\end{align*}
which have the form of the D4/D8/NS5 solution we started with. To obtain $AdS_4$ on the uncompactified space, one takes the near horizon limit $x^3\to 0$ which is equivalent to
\begin{equation*}
  H^\text{D8}\to\frac{2}{3} (x^3)^{-5/3}\,,\quad
  H^\text{D4}_i \to{2 \over 3 Q^4}\, \quad \text{and} \quad
  H^\text{NS5}_i \to Q (x^3)^{-2/3}\,.
\end{equation*}
In this limit
\begin{equation*}
  \d\tilde{s}_Q = \sum_{i=1}^6 (dy^i)^2
\end{equation*}
is the metric of a flat, 6-dimensional torus with
\begin{equation*}
  \beta^{31}=-\beta^{41}=Q y^6\quad\text{and}\quad
  -\beta^{32}=-\beta^{41}=Q y^5\,.
\end{equation*}

Having shown that the near horizon limit of four intersecting $Q$-branes leads to the metric of flat six-dimensional space, let us now
start from the Iwasawa manifold, the near horizon limit of four KK-monopoles, and perform two T-duality transformations along the  $y^3$ and $y^4$ directions of the internal manifolds.
The 6-dimensional metric takes the form
\begin{equation}\label{fourQmetric}
ds^2_{Q}={1\over x}\sum_{i=1}^4 (dy^i)^2 + \sum_{j=5,6} (dy^j)^2 
\end{equation}
where $x=1+Q^2\biggl((y^5)^2+(y^6)^2\biggr)$.
The corresponding $B$-field components are 
\begin{equation}\label{fourQB}
-B_{24}=B_{13} ={Qy^6\over x} \, ,\quad    B_{14}=B_{23}={Qy^5 \over x}\, .
\end{equation}
Since we are dealing with a non-geometric background, it is more convenient to re-express the metric $G$ by its redefined metric $\tilde G$ and $B$-field in terms
of the bi-vector $\beta$. Using eq.(\ref{eq:relation}), we obtain
\begin{equation}\label{fourQbeta}
\tilde G_{ij}=\delta_{ij}\, ,\qquad
\beta^{24}=-\beta^{13} ={Qy^6} \, ,\quad    -\beta^{14}=-\beta^{23}={Qy^5 }\, .
\end{equation}
Since  $\beta^{ij} $ is linear in $y^5$ reps. in $y^6$, it is easy to compute the corresponding $Q$-fluxes.
These will transform all four $f$-fluxes into the non-geometric fluxes
\begin{equation}\label{flux3}
Q^{24}_6=-Q^{13}_6=-Q^{14}_5=-Q^{23}_5=Q
\end{equation}
where by T-duality $Q\equiv f\equiv H$. 
Hence  all four $f$-fluxes are transformed into the non-geometric fluxes.

With these four $Q$-fluxes, the corresponding closed string background is non-commutative in the following directions:
\begin{eqnarray}
[Y^2(\tau,\sigma),Y^3(\tau,\sigma)] & \simeq & Q^{23}_5~ \tilde p^5\, ,\nonumber\\
\lbrack Y^1(\tau,\sigma),Y^4(\tau,\sigma)\rbrack & \simeq & Q^{14}_5~ \tilde p^5\, ,\nonumber\\
\lbrack Y^1(\tau,\sigma),Y^3(\tau,\sigma)\rbrack & \simeq & Q^{13}_6~ \tilde p^6\, ,\nonumber\\
\lbrack Y^2(\tau,\sigma),Y^4(\tau,\sigma)\rbrack & \simeq & Q^{24}_6~ \tilde p^6\, .
\end{eqnarray} 
The $Y^i(\tau,\sigma)$ are the internal closed string coordinates,  and the $\tilde p^5$ and $\tilde p^6$ are the dual (winding) momenta in the 5,6-directions of internal space.

Concerning the 4D low energy action, as already discussed in \cite{Shelton:2005cf}, the IIA superpotential with non-geometric $Q$-flux has the following general  structure:
\begin{eqnarray}
Q-{\rm flux:}\quad W_Q^{IIA}= a_{ij}ST_iT_j+ c_{ijm}T_iT_jU_m\, .
\end{eqnarray}
With our choice of four  $Q$-fluxes we then obtain
\begin{eqnarray}
 W_Q^{IIA}=Q^{24}_{6}ST_1T_2+Q^{23}_{5}T_1T_2U_1+Q^{14}_{5}T_1T_2U_2+Q^{13}_{6}T_1T_2U_3\, .
\end{eqnarray}

\subsubsection{Type IIA: one $H$-flux, two $f$-fluxes, one $Q$-flux}

We can consider also intersections among $Q$-branes and other (geometrical) brane configurations. Applying one T-duality along $y^3$  on the intersection configuration of two KK monopoles and
two NS 5-branes, we obtain the intersection of one NS 5-brane, two KK monopoles and one $Q$-brane:
\bigskip
\begin{center}
 \begin{tabular}{|c|c|c|c|c|c|c|}
    \hline
    & $y^1$  & $y^2$  & $y^3$  & $y^4$  & $y^5$  & $y^6$ \\
    \hline
     \hline
    $\mathrm{NS}5$ &   $\bigotimes$  &   &  $\bigotimes$ &
                 & $\bigotimes$ &  \\
\hline
    $\mathrm{KK}^{\prime}$    &  $\bigotimes$  &  &  $\bullet$ &  $\bigotimes$
                 &  & $\bigotimes$ \\
 \hline
    $\mathrm{Q}^{\prime\prime}$    & $\bullet$  & $\bigotimes$ & $\bullet$  &
               $\bigotimes$    &  $\bigotimes$  & \\
\hline
    $\mathrm{KK}^{\prime\prime\prime}$    & $\bullet$ & $\bigotimes$ &  $\bigotimes$     &
               &  &  $\bigotimes$ \\
\hline
   \end{tabular}
\end{center}
This is equivalent to one T-duality transformations along $y^3$ when starting from the IIB Nilmanifold $N_{5.1}$.
The resulting IIA background has the two geometric fluxes
\begin{equation}
f^3_{25}=-f^1_{45}=f\, .
\end{equation}
In addition the $H$-flux
\begin{equation}
H_{246}=H 
\end{equation}
survives the T-duality transformation. Finally,  the new flux, which we gain after the T-duality, is non-geometric $Q$-flux
\begin{equation}\label{flux2}
Q^{13}_6=-Q\, .
\end{equation}
The 6-dimensional metric takes the following form
\begin{equation}
  ds_{H,f,Q}^2 = \frac{1}{x}\biggl[ (dy^3 + f y^5 dy^2)^2 + (dy^1 - f y^5 dy^4)^2 \biggr] + \sum_{i=2,4,5,6} (dy^i)^2 
\end{equation}
where $x=1+Q^2(y^6)^2$.
The non-vanishing components of the $B$-field are 
\begin{equation}
  B_{12}=B_{34}={f x^5 Q x^6 \over x}\,,\quad  B_{13}={Q  y^6\over x}\quad \text{and} \quad B_{24}={\left(H + {(f x^5)^2 Q y^6 \over x}\right)y^6}\, .
\end{equation}

\vskip0.2cm

   \bigskip
The corresponding closed string background is non-commutative in the following directions:
\begin{eqnarray}
  \lbrack Y^1(\tau,\sigma),Y^3(\tau,\sigma)\rbrack \simeq Q^{13}_6~\tilde p^6\, .
\end{eqnarray} 

\vskip0.4cm

\subsubsection{Type IIB: two $f$-fluxes, two $Q$-fluxes}

We can also consider intersections of two KK monopoles and two $Q$-branes. They arise from the intersection of four KK monopoles after T-duality in $y^3$:

\vskip0.3cm
\begin{center} \begin{tabular}{|c|c|c|c|c|c|c|}
    \hline
    & $y^1$  & $y^2$  & $y^3$  & $y^4$  & $y^5$  & $y^6$ \\
    \hline
     \hline
       $\mathrm{KK}$ &   $\bigotimes$  & $\bullet$ &  $\bigotimes$ &
                 & $\bigotimes$ &  \\
\hline
    $\mathrm{Q}^{\prime}$    &  $\bigotimes$  & $\bullet$ & $\bullet$  &  $\bigotimes$
                 &  & $\bigotimes$ \\
 \hline
    $\mathrm{Q}^{\prime\prime}$    & $\bullet$ & $\bigotimes$ & $\bullet$  &
               $\bigotimes$    &  $\bigotimes$  & \\
\hline
    $\mathrm{KK}^{\prime\prime\prime}$    & $\bullet$ & $\bigotimes$ &  $\bigotimes$     &
               &  &  $\bigotimes$ \\
\hline
   \end{tabular}
   \end{center}

\vskip0.3cm

The near horizon geometry of this brane intersection can be obtained by direct computation, which are similar to the ones shown in section \ref{sec:fourQfluxes} . Alternatively, starting from the Iwasawa manifold, let us perform one T-duality transformation along the  $y^3$ direction.
This transformation will keep the following two $f$-fluxes untouched:
\begin{equation}
f^2_{46}=-f^1_{45}=f\,.
\end{equation}
In addition the remaining two $f$-fluxes are transformed  into the two non-geometric fluxes
\begin{equation}\label{qflux1}
Q^{13}_6=Q^{23}_5=Q\,.
\end{equation}
The 6-dimensional metric takes the following form
\begin{equation}
  ds^2_{f,Q}= (dy^1-f y^5 dy^4)^2 + (dy^2 +f y^6 dy^4)^2 + {(dy^3)^2 - \biggl(Q y^6 dy^1+Q y^5 dy^2\biggr)^2 \over x} + \sum_{i=4,5,6} (dy^i)^2
\end{equation}
where $x=1+Q^2\biggl((y^5)^2+(y^6)^2\biggr)$.
The non-vanishing components of the $B$-field are 
\begin{equation}
B_{13}={Qy^6\over x}\, ,\quad B_{23}={Qy^5\over x}\, .
\end{equation}

Let us again compute the background obtained after the field redefinition eq.(\ref{eq:relation}).
For the redefined metric we obtain
\begin{equation}
  d\tilde{s}^2_{f,Q}=\biggl(dy^4+f y^5 dy^1-f y^6 dy^2\biggr)^
  2+\sum_{i=1,2,3,5,6} (dy^i)^2
\end{equation}
and the bi-vector $\beta$ becomes in this case 
\begin{equation}
\beta^{13}=-{Qy^6}\, ,\quad \beta^{23}=-{Qy^5}\, .
\end{equation}
This is in good agreement with the $Q$-fluxes in eq.(\ref{qflux1}). The diffeomorphism
\begin{equation}
  \tilde{y^1}=y^4\,,\quad
  \tilde{y^2}=y^3\,,\quad
  \tilde{y^3}=-y^1\,,\quad
  \tilde{y^4}=y^2\,,\quad
  \tilde{y^5}=y^6\quad\text{and}\quad
  \tilde{y^6}=y^5
\end{equation}
combined with $H\equiv Q$ transforms $d\tilde{s}_{f,Q}$ and $\beta$ into the metric and the $H$-flux of the Nilmanifold $N_{5.1}$ presented in section \ref{nilmanifold N5.1}. Therefore, like $N_{5.1}$, the background described in this section exhibits a $SU(2)$ group structure.

\vskip0.5cm

This space is non-commutative in the following directions:
  \begin{eqnarray}
[Y^2(\tau,\sigma),Y^3(\tau,\sigma)] & \simeq & Q^{23}_5 ~\tilde p^5\, ,\nonumber\\
\lbrack Y^1(\tau,\sigma),Y^3(\tau,\sigma)\rbrack & \simeq & Q^{13}_6 ~\tilde p^6\, .
\end{eqnarray}

\subsubsection{Type IIA: four $R$-fluxes}

\vskip0.5cm
Now we are finally coming to 6D spaces with non-vanishing $R$-fluxes.
Starting from the flat torus with four $H$-fluxes, let us perform a T-duality in all six internal directions.
This results in the following intersection pattern of four $R$-branes:

\vskip0.5cm

\begin{center}
 \begin{tabular}{|c|c|c|c|c|c|c|}
    \hline
    & $y^1$  & $y^2$  & $y^3$  & $y^4$  & $y^5$  & $y^6$ \\
    \hline
     \hline
       $\mathrm{R}$ &   $\bigotimes$  & $\bullet$ &  $\bigotimes$ &$\bullet$
                 & $\bigotimes$ &  $\bullet$\\
\hline
    $\mathrm{R}^{\prime}$    &  $\bigotimes$  & $\bullet$ & $\bullet$  &  $\bigotimes$
                 & $\bullet$ & $\bigotimes$ \\
 \hline
    $\mathrm{R}^{\prime\prime}$    & $\bullet$ & $\bigotimes$ & $\bullet$  &
               $\bigotimes$    &  $\bigotimes$  &$\bullet$ \\
\hline
    $\mathrm{R}^{\prime\prime\prime}$    & $\bullet$ & $\bigotimes$ &  $\bigotimes$     &$\bullet$
               & $\bullet$ &  $\bigotimes$ \\
\hline
   \end{tabular}
   \end{center}
\vskip0.3cm
Using double geometry, the metric and the $B$-field of this configuration is obtained from the background of four intersecting $Q$-branes by replacing $y^5$ and $y^6$ by its dual coordinates $\tilde y^5$ and $\tilde y^6$. Therefore we simple obtain from eqs.(\ref{fourQmetric}), (\ref{fourQB}) and (\ref{fourQbeta}):
\begin{equation}\label{fourRmetric}
  ds^2_{R}={1\over \tilde{x}}\sum_{i=1}^4 (dy^i)^2 + \sum_{j=5,6} (d\tilde{y}^j)^2 
\end{equation}
where $\tilde x=1+R^2\biggl((\tilde y^5)^2+(\tilde y^6)^2\biggr)$.
The corresponding $B$-field components are 
\begin{equation}\label{fourRB}
-B_{24}=B_{13} ={R\tilde y^6\over \tilde x} \, ,\quad    B_{14}=B_{23}={R\tilde y^5 \over\tilde x}\, .
\end{equation}
and
\begin{equation}\label{fourRbeta}
\beta^{24}=-\beta^{13} ={R\tilde y^6} \, ,\quad    -\beta^{14}=-\beta^{23}={R\tilde y^5 }\, ,
\end{equation}
where by T-duality $R\equiv Q\equiv f\equiv H$.
The associated $R$-fluxes are
\begin{equation}
R^{246}=-R^{136}=-R^{145}=-R^{235}=R\,.
\end{equation}






One can easily work out the corresponding commutators and three-brackets. The phase space structure of the four $R$-brane intersection 
is quite interesting; e.g. for the 3-brackets among the closed string coordinates $Y^i(\tau,\sigma)$ one obtains
\begin{eqnarray}
\lbrack Y^2(\tau,\sigma),Y^4(\tau,\sigma),Y^6(\tau,\sigma)\rbrack &\simeq &R^{246}\, ,\nonumber\\
\lbrack Y^2(\tau,\sigma),Y^3(\tau,\sigma),Y^5(\tau,\sigma)\rbrack &\simeq &R^{235}\, ,\nonumber\\
\lbrack Y^1(\tau,\sigma),Y^3(\tau,\sigma),Y^6(\tau,\sigma)\rbrack &\simeq &R^{136}\, ,\nonumber \\
\lbrack Y^1(\tau,\sigma),Y^4(\tau,\sigma),Y^5(\tau,\sigma)\rbrack &\simeq &R^{145}\, .
\end{eqnarray}
The associated commutators also can be written down without big effort.

Finally, the IIA superpotential with non-geometric $R$-flux has the following general  structure:
\begin{eqnarray}
R-{\rm flux:}\quad W_R^{IIA}= a_{ijk}ST_iT_jT_k+ c_{ijkm}T_iT_jT_kU_m\, .
\end{eqnarray}
Here we obtain:
\begin{eqnarray}
 W_R^{IIA}=R^{246}ST_1T_2T_3+R^{235}T_1T_2T_3U_1+R^{145}T_1T_2T_3U_2+R^{136}T_1T_2T_3U_3\, .
\end{eqnarray}

\subsubsection{Type IIB: two $Q$-fluxes, two $R$-fluxes}

For completeness let us also perform a T-duality in the internal directions $y^1,y^2,y^3,y^4,y^5$, starting from from the flat torus with four $H$-fluxes.
This results in the following brane intersection pattern:

\vskip0.4cm

\begin{center} 
 \begin{tabular}{|c|c|c|c|c|c|c|}
    \hline
    & $y^1$  & $y^2$  & $y^3$  & $y^4$  & $y^5$  & $y^6$ \\
    \hline
     \hline
       $\mathrm{Q}$ &   $\bigotimes$  & $\bullet$ &  $\bigotimes$ &$\bullet$
                 & $\bigotimes$ &  \\
\hline
    $\mathrm{R}^{\prime}$    &  $\bigotimes$  & $\bullet$ & $\bullet$  &  $\bigotimes$
                 & $\bullet$ & $\bigotimes$ \\
 \hline
    $\mathrm{Q}^{\prime\prime}$    & $\bullet$ & $\bigotimes$ & $\bullet$  &
               $\bigotimes$    &  $\bigotimes$  & \\
\hline
    $\mathrm{R}^{\prime\prime\prime}$    & $\bullet$ & $\bigotimes$ &  $\bigotimes$     &$\bullet$
               & $\bullet$ &  $\bigotimes$ \\
\hline
   \end{tabular}
   \end{center}

\vskip0.3cm
The corresponding non-geometric fluxes are
\begin{equation}
Q^{24}_6=-Q^{13}_6=Q\, ,
\quad
-R^{145}=-R^{235}=R\, .
\end{equation}

This space is non-commutative/non-associative in the following directions:
  \begin{eqnarray}
\lbrack Y^1(\tau,\sigma),Y^3(\tau,\sigma)\rbrack & \simeq & Q^{13}_6 ~\tilde p^6\, \nonumber\\
\lbrack Y^2(\tau,\sigma),Y^4(\tau,\sigma)\rbrack & \simeq & Q^{24}_6 ~\tilde p^6\, \nonumber\\
\lbrack Y^2(\tau,\sigma),Y^3(\tau,\sigma),Y^5(\tau,\sigma)\rbrack &\simeq &R^{235}\, .
\end{eqnarray}

\section{Conclusions and summary}

In this paper we have discussed the construction of non-geometric $Q$- and $R$-branes as microscopic sources of non-geometric $Q$- and $R$-fluxes.
Having constructed these non-geometric branes via T-duality from the  NS 5-brane and the purely geometric KK-monopole background, 
we showed that doubled field theory and the use of the redefined background fields $\tilde G$ and $\beta$ leads to a very simple and elegant construction
of these brane solutions. This also allows for straightforward derivation of the associated $Q$- and $R$-background fields as certain derivatives of the
bi-vector $\beta$ background. Since the non-geometric branes carry an elementary unit of $Q$- and $R$-charge, it follows from the nontrivial monodromy properties
around these branes, that the transversal closed string geometry  is non-commutative or, respectively, non-associative.

In the second part of the paper we constructed supersymmetric type IIA and type IIB intersecting brane configurations, consisting of NS 5-branes, KK-monopoles, $Q$-
and $R$-branes. They lead to consistent six-dimensional compact space with various combinations of non-geometric fluxes and an interesting non-commutative or
non-associative geometric structure. 
We can summarize the web of T-dualities that led to all considered geometrical and non-geometrical brane intersection spaces as follows:
\vskip0.6cm
\begin{eqnarray}
\label{eq:TdualityChain2}
&~&\hskip4.1cm\stackrel{T_{3}}\longrightarrow \hskip0.2cm 1NS_5+2KK+1Q|_{IIA}\hskip0.2cm  \stackrel{T_{2}}\longrightarrow
\nonumber\\
&~&\hskip3.5cm \nearrow \hskip6.2cm  \searrow
\nonumber\\
&~&
4NS_5|_{IIA} \stackrel{T_{1}}{\longrightarrow}2NS_5+2KK|_{IIB}
\stackrel{T_{2}}{\longrightarrow}  4KK|_{IIA}
\stackrel{T_{3}}
{\longrightarrow}
2KK+2Q|_{IIB}
\stackrel{T_{4}}{\longrightarrow} 4Q|_{IIA}\nonumber\\
&~&\hskip12.2cm  \stackrel{T_5}{\downarrow} \nonumber\\
&~&\hskip11.2 cm  2Q+2R|_{IIB}\nonumber\\
&~&\hskip12.2 cm \stackrel{T_6}{\downarrow}
\nonumber\\
&~&\hskip12 cm 4R|_{IIA}
\nonumber
\end{eqnarray}

\vskip0.5cm

\noindent
In the corresponding effective four-dimensional theories, all moduli of the associated 6D compact spaces are fixed by the combinations of geometric and/or
non-geometric fluxes, when also including Ramond fluxes from additional intersecting D-branes.
The corresponding potential has in its minimum a negative energy, confirming the fact that in the near horizon limit of the interesting branes the four-dimensional space
is given by $AdS_4$.

\vskip2cm

\subsection*{Acknowledgments}

This work is supported by the Munich Excellence Cluster for Fundamental Physics "Origin and Structure of the Universe".
We like to thank D. Andriot, A. Betz, O. Hohm, M. Larfors, D. Tsimpis and T. Weigand for very useful discussions and comments on the manuscript.
D.L. likes to thank the theory group of CERN and also the MIT Center for Theoretical Physics for hospitality, where part of this work was performed.

\newpage
\appendix
\section{$Q$-branes as solution of the NS action}\label{Qbrane eom}
Setting the variation $\delta S$ of the standard NS action in eq.(\ref{standardNS}) to zero, leads to the field equations 
\begin{eqnarray}
  && \R + 4(\nabla^2\phi - (\del\phi)^2) = \frac{1}{12} H_{ijk}H^{ijk}\label{eom standard dilaton}\\
  && \R_{ij} + 2\nabla_i\del_j\phi = \frac{1}{4} H_{imn}H_j^{\phantom{j}mn}\label{eom standard metric} \\
  && 2\del^i\phi H_{ijk} = \nabla^i H_{ijk}\label{eom standard B}
\end{eqnarray}
for the metric $G_{ij}$, the asymmetric Kalb-Ramond field $B_{ij}$ with the $H$-flux $H_{ijk}=\nabla_{[i}B_{jk]}$ and the dilaton $\phi$. When evaluating the various terms of these field equations, it is sufficient to consider only the orthogonal coordinates $x^3$, $y^1$, $y^2$ and $y^3$ of the $Q$-brane. The remaining, transversal coordinates are trivial and do not give a contribution. Using eqs.(\ref{qbrane}) and (\ref{BQ}), we start with
\begin{equation}
G=\begin{pmatrix}
  h & 0 & 0 & 0 \\
  0 & h \over h^2+A_2^2 & 0 & 0 \\
  0 & 0 & h \over h^2+A_2^2 & 0 \\
  0 & 0 & 0 & h \\
\end{pmatrix}\, \text{,} \quad
B=\begin{pmatrix}
  0 & 0 & 0 & 0 \\
  0 & 0 & A \over h^2+A_2^2 & 0 \\
  0 & - {A \over h^2+A_2^2 }& 0 & 0\\
  0 & 0 & 0 & 0 \\
\end{pmatrix} \quad \text{and} \quad
\phi = \log\sqrt{h \over h^2+A_2^2}\nn
\end{equation}
where $h(x^3,y^3)$ is an arbitrary harmonic function and $A_2(x^3,y^3)$ is linked to $h$ by eq.(\ref{A2 condition}). This two functions $h$ and $A_2$ parameterize the $Q$-brane completely. They fulfill the Cauchy-Riemann equations
\begin{equation}
  h^{(1,0)}= A_2^{(0,1)}\quad\text{and}\quad
  h^{(0,1)}=-A_2^{(1,0)}\nn\,,
\end{equation}
where we used the convenient abbreviation $h^{(n,m)}=\del_{x^3}^n\del_{y^3}^m h$. This property of $h$ and $A_2$ allows to express the $Q$-brane as an hermitian manifold which is parameterized by a single holomorphic function
\begin{equation*}
  f(z^2) = h + i A_2 \quad \text{with} \quad z^2=x^3 + i y^3\,.
\end{equation*}
The metric on this hermitian manifold 
\begin{equation}
  ds^2_\text{h}={1\over f(z^2)}dz^1\otimes d\overline{z}^1 +
    f(z^2) dz^2\otimes d\overline{z}^2
\end{equation}
reproduces for $z_2=y^1 + i y^2$ to the metric $ds^2=1/2(ds^2_\text{h}+\overline{ds}^2_\text{h})$ corresponding to $G_{ij}$ and the 2-form flux $B=i/2(ds^2_\text{h}-\overline{ds}^2_\text{h})=B_{ij}dy^i\wedge dy^j$.

To simply the following calculation, we use the substitution rules
\begin{equation}
A^{(0,1)} \to  h^{(1,0)}\,\text{,}\quad
A^{(1,0)} \to -h^{(0,1)}\,\text{,}\quad
A^{(2,0)} \to -h^{(1,1)}\quad\text{and}\quad
A^{(0,2)} \to  h^{(1,1)}\nn
\end{equation}
in order to eliminate all $A_2$-derivatives from the field equations. Additionally the substitution $h^{(0,2)}\to -h^{(2,0)}$ takes into account that $h$ is a harmonic function. Using the Mathematica tensor package xAct, we computed all terms of the field equations. The terms
\begin{eqnarray}
  && \R=3 {{h^{(1,0)}}^2+{h^{(0,1)}}^2 \over 2 h^3} \,\text{,}\quad
  \nabla^2\phi=0 \,\text{,}\quad
  (\del\phi)^2={{h^{(1,0)}}^2+{h^{(1,0)}}^2 \over 2 h^3} \,\text{,}\nn\\
  && H_{ijk} H^{ijk} = 6{{h^{(1,0)}}^2+{h^{(0,1)}}^2 \over h^3}\,\text{,}\quad
  (\del^i\phi) H_{ijk}=0\quad\text{and}\quad
  \nabla^i H_{ijk}=0\nn
\end{eqnarray}
of eqs.(\ref{eom standard dilaton}) and (\ref{eom standard B}) can be reduced to a simple form which only depends on $h$ and its first derivative. This is not the case for terms of eq.(\ref{eom standard metric}) which are much more complex and therefore not presented here. Nevertheless they fulfill, like the terms given above, the field equation derived from the standard NS action.

Now we like to discuss the solutions of the redefined NS action in eq.(\ref{Qaction}).
The principle of stationary action $\delta\tilde S=0$ leads for the redefined NS action to the following redefined field equations \cite{Andriot:2011uh}
\begin{eqnarray}
  && \tilde\R + 4(\nabla^2\tilde\phi - (\del\tilde\phi)^2) = \frac{1}{4} Q_i^{\phantom{i}jk}Q^i_{\phantom{i}jk}\label{eom redef dilaton}\\
  && \tilde\R_{ij} + 2\nabla_i\del_j\tilde\phi = \frac{1}{4} \left(Q_{i}^{\phantom{i}mn}Q_{jmn}-2 Q_{mi}^{\phantom{mi}n}Q^{m}_{\phantom{m}jn}\right)\label{eom redef metric} \\
  && 2(\del^i\tilde\phi) Q_i^{\phantom{i}jk} = \nabla^i Q_i^{\phantom{i}jk}\,\text{.}\label{eom redef B}
\end{eqnarray}
where $Q$ denotes the $Q$-flux $Q_{i}^{\phantom{i}jk}=\del_i\beta^{jk}$. Like for the field equations of the standard NS action, we only consider the orthogonal coordinates of the $Q$-brane whose configuration
\begin{equation}
\tilde{G}=\begin{pmatrix}
  h & 0 & 0 & 0 \\
  0 & 1 \over h & 0 & 0 \\
  0 & 0 & 1 \over h & 0 \\
  0 & 0 & 0 & h \\
\end{pmatrix}\, \text{,} \quad
\beta=\begin{pmatrix}
  0 & 0 & 0 & 0 \\
  0 & 0 & -A & 0 \\
  0 & A & 0 & 0\\
  0 & 0 & 0 & 0 \\
\end{pmatrix} \quad \text{and} \quad
\tilde\phi = \log\sqrt{1 \over h}\nn
\end{equation}
is given by eq.(\ref{Qbranered}). Again the above introduced substitutions are applied during the calculations of the individual terms of the redefined field equations, leading to
\begin{eqnarray}
  && \tilde\R=-5 {{h^{(1,0)}}^2+{h^{(0,1)}}^2 \over 2 h^3} \,\text{,}\quad
  \nabla^2\tilde\phi={{h^{(1,0)}}^2+{h^{(0,1)}}^2 \over h^3} \,\text{,}\quad
  (\del\tilde\phi)^2={{h^{(1,0)}}^2+{h^{(1,0)}}^2 \over 4 h^3} \,\text{,}\nn\\
  && Q_i^{\phantom{i}jk}Q^i_{\phantom{i}jk}=2{{h^{(1,0)}}^2+{h^{(0,1)}}^2 \over h^3}\,\text{,}\quad
  (\del^i\tilde\phi) Q_i^{\phantom{i}jk}=0\quad\text{and}\quad
  \nabla^\mu Q_i^{\phantom{i}jk}=0\nn
\end{eqnarray}
for eqs.(\ref{eom redef dilaton}) and (\ref{eom redef B}). Additionally on get the following terms
\begin{eqnarray}
\tilde\R_{ij}=
\begin{pmatrix}
 \frac{2 {h^{(0,1)}}^2-3 {h^{(1,0)}}^2+2 h h^{(2,0)}}{2 h^2} & 0 & 0 &  \frac{2 h h^{(1,1)}-5 h^{(0,1)} h^{(1,0)}}{2
   h^2} \\
 0 & -\frac{{h^{(0,1)}}^2+{h^{(1,0)}}^2}{h^4} & 0 & 0 \\
 0 & 0 & -\frac{{h^{(0,1)}}^2+{h^{(1,0)}}^2}{h^4} & 0 \\
 \frac{2 h h^{(1,1)}-5 h^{(0,1)} h^{(1,0)}}{2 h^2} & 0 & 0 & -\frac{3 {h^{(0,1)}}^2-2{h^{(1,0)}}^2+2 h h^{(2,0)}}{2 h^2}
\end{pmatrix}\nn && \nn \\
\nabla_i\del_j\tilde\phi =
\begin{pmatrix}
 -\frac{{h^{(0,1)}}^2-3 {h^{(1,0)}}^2+2 h h^{(2,0)}}{4 h^2} & 0 & 0 & \frac{2 h^{(0,1)} h^{(1,0)}-h h^{(1,1)}}{2 h^2} \\
 0 & \frac{{h^{(0,1)}}^2+{h^{(1,0)}}^2}{4 h^4} & 0 & 0 \\
 0 & 0 & \frac{{h^{(0,1)}}^2+{h^{(1,0)}}^2}{4 h^4} & 0 \\
 \frac{2 h^{(0,1)} h^{(1,0)}-h h^{(1,1)}}{2 h^2} & 0 & 0 & \frac{3 {h^{(0,1)}}^2-{h^{(1,0)}}^2+2 h h^{(2,0)}}{4 h^2}
\end{pmatrix} && \nn \\
Q_{i}^{\phantom{i}mn}Q_{jmn}-2 Q_{mi}^{\phantom{mi}n}Q^{m}_{\phantom{m}jn} =
\begin{pmatrix}
 \frac{2 {h^{(0,1)}}^2}{h^2} & 0 & 0 & -\frac{2 h^{(0,1)} h^{(1,0)}}{h^2} \\
 0 & -2 \frac{{h^{(0,1)}}^2+{h^{(1,0)}}^2}{h^4} & 0 & 0 \\
 0 & 0 & -2 \frac{{h^{(0,1)}}^2+{h^{(1,0)}}^2}{h^4} & 0 \\
 -\frac{2 h^{(0,1)} h^{(1,0)}}{h^2} & 0 & 0 & \frac{2 {h^{(1,0)}}^2}{h^2}
\end{pmatrix}\nn
\end{eqnarray}
for eq.(\ref{eom redef metric}). As expected, these results prove that the redefined $Q$-brane satisfies the redefined field equations.

\newpage

\providecommand{\href}[2]{#2}\begingroup\raggedright\endgroup

\end{document}